\documentclass[preprint,12pt,3p]{elsarticle}
\usepackage{enumitem}
\usepackage{multicol}
\usepackage{amssymb}
\usepackage{xcolor}
\usepackage{sectsty}
\usepackage{amsmath}
\usepackage{psfrag}
\usepackage{caption}
\usepackage{microtype}
\usepackage{subfig}
\usepackage{graphics}
\usepackage{color}
\usepackage{tikz}
\usepackage{url}
\usepackage{numcompress}
\DeclareGraphicsExtensions{.pdf,.png,.jpg}
\newcommand*{\boxedcolor}{red}
\makeatletter
\allowdisplaybreaks
\renewcommand{\boxed}[1]{\textcolor{\boxedcolor}{%
		\fbox{\normalcolor\m@th$\displaystyle#1$}}}
\makeatother
\newcommand{\BEAS}{\begin{eqnarray*}}
\newcommand{\EEAS}{\end{eqnarray*}}
\newcommand{\BEQ}{\begin{equation}}
\newcommand{\EEQ}{\end{equation}}
\newcommand{\BIT}{\begin{itemize}}
\newcommand{\EIT}{\end{itemize}}

\newcommand{\argmin}{\mathop{\rm argmin}}

\journal{Elsevier Science Direct Power System Research Journal}
\begin{document}
	\begin{frontmatter}
		\title{Look-Ahead SCOPF (LASCOPF) for Tracking Demand Variation via Auxiliary Proximal Message Passing (APMP) Algorithm
			\tnoteref{label0}}
		\tnotetext[label0]{The objective of this paper is to present novel algorithmic and computational tools for solving the Look-Ahead Security Constrained Optimal Power Flow (LASCOPF) accurately, provably, and reasonably fast. It will form a foundation, based on which, we will build upon further improvements of the algorithm and also the problem. The authors were supported, in part, by the National Science Foundation (NSF) under grant ECCS-1406894.}
		\author[label1]{Sambuddha Chakrabarti\corref{cor1}}
		\address[label1]{Department of Electrical \& Computer Engineering (ECE), The University of Texas at Austin,
			Austin, TX 78705, USA}
		\cortext[cor1]{I am the corresponding author}
		\ead{schakrabarti@utexas.edu}
		\author[label1]{Ross Baldick}
		\ead{baldick@ece.utexas.edu}
		\begin{abstract}
			In this paper, we will consider the Look-Ahead Security Constrained Optimal Power Flow (LASCOPF) problem looking forward multiple dispatch intervals, in which the load demand varies over dispatch intervals according to some forecast. We will consider the base-case and several contingency scenarios in the \emph{upcoming} as well as in the \emph{subsequent} dispatch intervals. We will formulate and solve the problem in a Model Predictive Control (MPC) paradigm. We will present the \emph{Auxiliary Proximal Message Passing (APMP)} algorithm to solve this problem, which is a bi-layered decomposition-coordination type distributed algorithm, consisting of an outer \emph{Auxiliary Problem Principle (APP)} layer and an inner \emph{Proximal Message Passing (PMP)} layer. The \emph{APP} part of the algorithm distributes the computation across several dispatch intervals and the \emph{PMP} part performs the distributed computation within each of the dispatch interval across different devices (i.e. generators, transmission lines, loads) and nodes or nets. We will demonstrate the effectiveness of our method with a series of numerical simulations.\\  
		\end{abstract}
	\end{frontmatter}
	\section{Introduction}
	%
	%
	%
	%
	\noindent In this paper, we will consider the $(N-1)$ \emph{Security-Constrained Optimal Power Flow (SCOPF)} problem, in which devices (i.e. generators, transmission lines, loads etc.) are connected on the power network and there exists a set of \emph{scenarios} --- each corresponding to the failure of a particular transmission line --- over which we must ensure optimal feasible operation of the network. But, instead of solving the SCOPF for the upcoming dispatch interval only, at the start of each such interval, we will also look forward several subsequent dispatch intervals, in which the load demands change according to a forecast. We will take into account the ramp rates of generators and operate the system such that the generation cost is minimized, subject to the constraints of satisfying the changing load demand, line power flow limit constraints, minimum and maximum generation constraint of each generator, enforced in each of the upcoming and subsequent dispatch intervals considered.
	
	The goal is to minimize a composite cost function that includes the cost (and constraints) of nominal operation, as well as those associated with constraints on operation in any of the (adverse) scenarios, for multiple dispatch intervals. This results in a large optimization problem, since variables in the network, namely, real power injection and bus voltage phase angle, are repeated $|\mathcal{L}|$ times for each of the $|\Omega|$ dispatch intervals, where $|\mathcal{L}|$ is the number of contingencies. Therefore we present a distributed algorithm to solve this problem, whereby we break the problem into several smaller independent sub-problems and solve them in parallel, such that the only co-ordination required is local, through the Locational Marginal Prices (LMP) updates. Specifically, we use a suitably modified version of the message passing algorithm from~\cite{KC:13} and~\cite{CK:14} to solve this problem efficiently.
	
	For simplicity, we consider only DC power flow in this paper.  The extension to AC power flow, involves simply applying the AC-OPF model from \cite{BWW:08}, \cite{LL:12}, \cite{SL:12a}, \cite{SL:12b}, \cite{LTZ:12}, and \cite{E:14} to each scenario and requiring that the phase angles of a given device are equal across all scenarios in the respective time periods. 
	
	In the previous works in \cite{KC:13}, \cite{LKWZ:13}, and \cite{CK:14}, Kraning \emph{et al.}, applied the Proximal Message Passing Algorithm to solve the standard Static Optimal Power Flow (OPF) Problem while Liu \emph{et al.} and Chakrabarti \emph{et al.} respectively applied it to solve the $(N-1)$ Security Constrained OPF (SCOPF). In this paper, we extend this approach to solving a look-ahead dispatch problem, which considers the variation of load over a time horizon. There can be interesting and direct underpinnings of solving such problems when it comes to design and allocation of Financial Transmission Rights (FTRs) under random outages, as well \cite{SamMo2018}. Our formulation for the LASCOPF is based on the Model Predictive Control (MPC) or Receding Horizon Control (RHC) methodology \cite{QB:03,CTHK:03,Her:05,TR:04,KW:11}. The rest of the paper is organized as follows: In section \ref{literature} we present a brief literature survey, after which, in section \ref{conventional} we introduce the system of notations and present the conventional formulation of look-ahead SCOPF in the ``Angles Represented" version, where the different real power flows on the Transmission Lines are represented in terms of the real powers injected at the buses and the voltage phase angles at the buses. In section \ref{APMPAlgo}, we apply the coarse-grained APP decomposition to split the problem into different dispatch intervals. In section \ref{DTNFormulation}, we reformulate all the preceding SCOPF problems (belonging to different dispatch intervals) into a different framework, the $\mathcal{DTN}$ (Devices-Terminals-Nets) formulation, which is particularly suitable for applying the ADMM (Alternating Direction Method of Multipliers) \cite{BP:11} based Proximal Message Passing algorithm to the problems. In section \ref{PMPAlgorithm}, we present the Proximal Message Passing fine-grained decomposition for the present problem. In section \ref{results}, we present the results of some simulation studies conducted on the IEEE test systems and we draw the concluding remarks and point to future research in section \ref{conclusion}.

	
	\section{Literature Survey and Related work}
	\label{literature}
	\noindent The Optimal Power Flow Problem is at the heart of Power Systems planning and operations. Initially formulated by Carpentier in \cite{Carp:62}, it has been studied for more than half a century, and has extensively been applied in the industry as well, including solving the electricity market scheduling and dispatch calculations by the ISOs/TSOs (which involves solving the DC-OPF) and applications by the market participants for planning using production-cost modeling systems, like Uplan \cite{Sambuddha2010}. Happ in \cite{Happ:77} and Chowdhury \& Rahman in \cite{CR:90} trace the historical development of OPF upto the 90s. A recent reference that provides a good summary of the historical development of the problem is Cain, O'Neill and Castillo \cite{cain2012history}. The references cited there also provide good insights into formulation and modeling particularly, of the ACOPF. There has been recently a surge in the interest in ACOPF following the works of Bai \emph{et al.} in \cite{BWW:08} and Lavaei \emph{et al.} in \cite{LL:12}, \cite{SL:12a}, \cite{SL:12b}, \cite{LTZ:12}, where the authors showed that the ACOPF problem can be solved accurately by a convex relaxation of the original problem into a Semidefinite Programming problem. In reference \cite{farivar2013branch}, the authors have presented the AC-OPF convex relaxations using Second Order Cone Programming (SOCP) for branch flow models, while reference \cite{low2014convex} by Low provides an excellent and comprehensive guide to the current state of research in AC-OPF convex relaxation, making a special mention of the theory of chordal graphs, clique tree-decomposition, and SDP, as applied to power networks. Utlilizing these two ideas, recent references \cite{dall2013distributed} and \cite{zheng2015fully} mention applying distributed algorithms to AC-OPF, respectively, in the contexts of a distribution level microgrid and reactive power compensation. Reference \cite{ADMMRamMadJav} illustrates application of ADMM to sparse SDP for solving AC-OPF. In the recent reference \cite{MHANNA201891}, the authors presented a modified version of the ADMM, showing that a global optimum for the highly non-convex AC-OPF can be obtained by applying this algorithm. The recent trend of applying distributed optimization algorithms to power flow problems in order to speed up the computation time has gained ample momentum already. Two of the recent papers that provide very good and comprehensive comparisons between the performance of several different algorithms for solving the OPF and SCOPF problems are \cite{SurvMol} and \cite{SurvKarg}, where the authors mentioned ADMM, Proximal Message passing (PMP), APP, Consensus+Innovation (C+I), Optimality Condition Decomposition (OCD) etc. algorithms.
	
	The pioneering work on the Security Constrained OPF (SCOPF) was done by Stott \emph{et al} in \cite{1458198}. Some of the recent references include works by Chiang \emph{et al.} \cite{CG:11}, Phan \emph{et al.} \cite{PK:12}, Chakrabarti \emph{et al.} \cite{CK:14} etc. The last two deal with application of distributed algorithms to the SCOPF problem. In \cite{E:14} and in \cite{LKWZ:13}, the authors have applied ADMM to solve the ACOPF and SCOPF, but not to a look-ahead problem. Significant early works on ADMM method during the 70s and 80s were \cite{GlM:75}, \cite{Gab:83}, \cite{FG:83} etc. followed by work during the 90's which include \cite{EF:93}, \cite{Eck:94b}. Combining these two fields gives rise to the Distributed Computational methods for OPF \cite{KB:97}, \cite{KB:00}. The last reference provides a good comparison of the distributed methods till the end of 90s. Good references on Model Predictive Control (MPC) or Receding Horizon Control (RHC) are \cite{QB:03,CTHK:03,Her:05,TR:04,KW:11}. \subsection{Contribution of the Present Work}While there has been some excellent work in the recent past on $(N-1)$ and $(N-k)$ SCOPFs as evidenced by references \cite{capitanescu2011state, lubin2015robust, amjady2011security, la1998line, articleSCOPF}, there is a dearth of literature on multiple-dispatch interval SCOPF. Also, a vast majority of the academic literature present till date on the topic of SCOPF either solve a corrective control version of the SCOPF (by changing generator dispatch to relieve line congestion in the outaged cases) or they solve a chance constrained version. Moreover, in the academic body of literature, there is also a severe lack of materials describing the software implementation for SCOPF and LASCOPF for the preventive control version. The main contribution of our work, in the light of the above, can be summarized as presenting the preventive control version of $(N-1)$ SCOPF for multiple dispatch intervals, as a part of which, we have also developed a software for implementing the mathematical model and the algorithms, which utilizes a two-layered distributed algorithm. In our present work, we will be solving the Look-Ahead Security Constrained Optimal Power Flow (LASCOPF) problem, formulated in a MPC/RHC format, using a combination of APP and ADMM-PMP algorithms, which we will call the APMP algorithm. As in all MPC/RHC formulations, the implicit assumption will be that this problem for several look-ahead dispatch intervals is solved at the beginning of each dispatch interval, updating the forecasts, thereby reaching the best accuracy and ability to track the time-varying load demand.
	\subsection{Comparison of the Present Algorithm Against Other Methods}In the present method, that we are about to describe in this paper, the heart of the problem lies in solving the OPF problem using ADMM-PMP, while using APP to reach consensus among multiple different OPFs that we solve. There are other methods also to achieve the same goal. It's hard to say, which method is superior to which one. There are many factors and considerations, and the merit of a particular method depends on what factors are considered important, for a particular application. In the light of that, in the recent papers, \cite{SurvKarg} and \cite{SurvMol} we have made comparisons against several different methods, which can be summarized as below:\\
	The different algorithms compared are Analytical Target Cascading (ATC), ADMM (the pristine version), ADMM-PMP, APP, Optimality Condition Decomposition (OCD), and Consensus+Innovation (C+I) in the light of the following factors:\\
	\begin{itemize}
		\item \textit{Presence of a central coordinator:} While ATC and ADMM do need central coordinators, ADMM-PMP, APP, OCD, and C+I do not need a central coordinator. This can speed up computations by reducing bottleneck of communication time with coordinator.
		\item \textit{Computations Effort:} For a single iteration, ADMM-PMP and C+I has got much less computational effort as compared to all the other algorithms. However, ADMM-PMP needs many more iterations to converge. In our case, we compromised on accuracy, in order to cut down on the ADMM-PMP iterations. In order to solve the LASCOPF, since we need to wrap the ADMM-PMP layer with the APP outer layers, we cannot help reduce the net computational burden. However, it is considerably less than had any of the other methods been used. 
		\item \textit{Data Exchanged and Shared: }ADMM-PMP has a high volume of data exchanged and shared as compared to the other algorithms. However, given the previous point, combined with the fact that the data shared and exchanged volume of APP is quite low, the net communication time of APMP is moderate.
	\end{itemize}
	From the above comparison, we can conclude that APMP is a good candidate for distributed algorithm for solving LASCOPF problems. Certainly, for big networks and many contingency scenarios and dispatch intervals, we wouldn't prefer a centralized algorithm. Given below, is a table (Table \ref{table:APMPCentComp}) indicating comparison of APMP against a centralized solver implemented with GUROBI, as a benchmark solver. The percentage difference in optimal objective value is calculated as:\\
	$\%\text{ Difference}=\frac{(\text{APMP Optimal Objective}-\text{Centralized Optimal Objective})\times 100}{\text{Centralized Optimal Objective}}$
	\begin{table}[ht] 
		
		\caption{Comparison of APMP against Centralized benchmark.} 
		
		\centering 
		
		\begin{tabular}{| c | c |} 
			
			\hline\hline 
			
			Type of the System & Percentage Difference \\ [0.5ex] 
			
			
			\hline 
			
			5 bus &	0 \\ [0.5ex] 
			\hline
			14 bus & 0.101128791 \\ [0.5ex] 
			\hline
			30 bus &	0 \\ [0.5ex] 
			\hline
			57 bus &	-0.003811375 \\ [0.5ex] 
			\hline
			118 bus &	0 \\ [0.5ex] 
			\hline
			300 bus &	-0.0001387213 \\ [0.5ex] 
			\hline
		\end{tabular} 
		
		\label{table:APMPCentComp} 
		
	\end{table}
	\section{Conventional Look-Ahead SCOPF Formulation}
	\label{conventional}
	In this section, we will formulate the models for the look-ahead SCOPF in the conventional framework. First we will introduce the notations and symbols to be used for the rest of the paper, which are same as those introduced in \cite{CK:14} with a few additions.
	\subsection{Notations and Conventions}
	\label{notations}
	We have categorized the entities used in the subsequent formulations into five different groups: Sets, Elements, Index, Parameters, and Variables.\\
	\begin{itemize}
		\item \textit{Sets}
		\begin{itemize}[label={}]
			\item $\mathcal{D}$: Set of Devices
			The next three sets form partitions of the set of devices:
			\item $G\subseteq\mathcal{D}$: Set of Generators
			\item $\mathbb{T}\subseteq\mathcal{D}$: Set of Transmission Lines
			\item $L\subseteq\mathcal{D}$: Set of Loads
			\item $\mathcal{T}$: Set of Terminals
			\item $\mathcal{N}$: Set of Nets or Buses or Nodes
			\item $J(N_i)$: Set of buses/nodes/nets directly connected to node $N_i\in\mathcal{N}$
			\item $\mathcal{L}=\{0,1,2,...,|\mathcal{L}|\}$: Set of possible $(N-1)$ Contingencies. The element 0 indicates the base case
			\item $\Omega=\{0, 1, 2, ..., |\Omega|\}$: Set of Dispatch intervals or, the net Dispatch Horizon under consideration. 1 indicates the upcoming dispatch interval under consideration, for which a dispatch decision will definitely be implemented. The rest, $\{2, 3, ..., |\Omega|\}$, are subsequent intervals in future. Hence, dispatch interval $0$  is the current running one. 
			\item $\dagger$ will be used to denote the transpose of a vector or matrix.
		\end{itemize}
		\item \textit{Elements}
		\begin{itemize}[label={}]
			\item $t$: Elements of $\mathcal{T}$
			\item $g$: Elements of $G$
			\item $D$: Elements of $L$
			\item $T$: Elements of $\mathbb{T}$
			\item $N$: Elements of $\mathcal{N}$
		\end{itemize}
		\item \textit{Indices}
		\begin{itemize}[label={}]
			\item $i,j$: Nets
			\item $k$: Terminals
			\item $q$: Generators
			\item $r$: Transmission Lines
			\item $d$: Loads
			\item $c$: Contingencies
			\item $\tau$: Dispatch Intervals
			\item $\nu$: Iteration count for ADMM/PMP algorithm
			\item $\mu_{APP}$: Iteration count for APP algorithm
		\end{itemize}
		\item \textit{Parameters}
		\begin{itemize}[label={}]
			\item $R_{T_r}, X_{T_r}, Z_{T_r}=R_{T_r}+(\sqrt{-1})X_{T_r},  B_{T_r}$: Resistance, Reactance, Impedance, and Susceptance of the $r^{th}$ Transmission Line, respectively
			\item $\alpha$, $\beta$, $\gamma$: APP tuning parameters, $\rho$: ADMM-PMP tuning parameter
			\item $\alpha_{g_q}, \beta_{g_q}, \gamma_{g_q}$: Quadratic, Linear, and Constant Cost Co-efficients of the $q^{th}$ Generator
			\item $C_{g_q}(.), f_{dev}(.)$ will be used to denote the cost function of the ${g_q}^{th}$ generator and that of a generic device, respectively, throughout. We will introduce the other cost functions in the appropriate sections
			\item $I_{\leq}(.)$, $I_{\geq}(.)$, $ I_{=}(.)$: Indicator functions such that $I_{\leq}(x)=0$, if $0\leq x$, and $=\infty$ otherwise, and similarly for $I_{\geq}(.)$ and $ I_{=}(.)$ 
			\item $\overline{P}_{g_q},\underline{P}_{g_q}, \overline{R}_{g_q},\underline{R}_{g_q}$(=$-\overline{R}_{g_q}$,usually), $\overline{L}_{T_r}$ denote the maximum and minimum generating limits of generators, maximum ramp-up and ramp-down limits of generators, and power carrying capacity of transmission lines respectively
		\end{itemize}
		\item \textit{Variables}
		\begin{itemize}[label={}]
			\item $P$: The real power or, MW associated with different devices
			\item $\theta$: The bus voltage phase angles associated with different nodes, as well as the devices
			\item $\lambda$: The lagrange multiplier or dual variable corresponding to the consensus among different power generation variables between the dispatch intervals
		\end{itemize}
	\end{itemize}
	The following is the convention we follow in order to identify the associations of any particular variable to the sets:\\ $\Big(x_{d(p_1)(p_2)}^{(p_3)(p_4)}\Big)^{(\nu)}$.\\In the above, $x_d$ is a variable associated with a device-terminal or node/net-terminal, $d$. ${(p_1)(p_2)}$ and ${(p_3)(p_4)}$ indicate the computational units/agents, denoted, respectively, by a combination of $p_1, p_2$ and $p_3, p_4$, where a particular $p_i$ may refer to either a dispatch interval or contingency scenario it is associated with. The combination ${(p_1)(p_2)}$ denotes a particular computational agent linked to this particular combination. The above notation should be interpreted as the belief about the value of the variable, $x_d$ associated or linked to the superscripted computational agent (${(p_3)(p_4)}$), as held by the subscripted computational agent (${(p_1)(p_2)}$) at the $\nu^{-\text{th}}$ iteration. Let's take a look at the next example, to make the meaning clearer.\\ $\Big(P_{{N}_{t}(c_1)(\tau_1)}^{(c_2)(\tau_2)}\Big)^{(\nu)}$.\\The foregoing notation refers to the belief of the value of a variable $P_{N_t}$ associated with a particular terminal, $N_t$, (which is indexed by the $t$) of either a net or a device for the contingency scenario indexed by $c_2$ during $\tau_2$, as estimated by a computational agent linked to $\tau_1$ and contingency scenario, $c_1$. Sometimes we will use the net number instead of the terminal number in the above convention, when we want to indicate several devices connected to a particular net. If it is part of an iterative algorithm, then the outermost superscript $\nu$ indicates the iteration count. Whenever a variable is boldface, one or more of the indices will be missing and that means the boldface variable is a vector each of whose components will have all or some of the missing indices (the components themselves can be vectors or scalars). When the variable is not bold-face and still some of the indices are missing, that means it is a scalar and the missing indices are either irrelevant or their values are implied from the context. Also it is to be observed that since generators and loads are single terminal devices, it is not necessary to specify the terminals for these, unless absolutely required.
	\subsection{Look-Ahead Dispatch: Generalized Case of Demand Variation for Multi-Bus Case}
	In this paper we build on the static SCOPF problem formulation in \cite{CK:14}. We will develop the mathematical model for a look-ahead dispatch calculation that considers an upcoming and several subsequent dispatch intervals at the onset of each current dispatch interval and takes into account the possible variations of operating parameters across different scenarios represented in different subsequent dispatch intervals, so that at each interval the entire system is secure. At the end of each current dispatch interval, the calculation ``rolls forward" and the whole look-ahead calculation is repeated with updated demand forecast, just as in a ``Model Predictive Control (MPC)" or a ``Receding Horizon Control (RHC)." The generalized version of this problem for a multi-time horizon, arbitrary network is as follows:
	\begin{subequations}\label{Look-Ahead_General_Ang_Inc}
		\begin{gather}
		\min_{P_{g_q}^{(\tau)},\theta}\sum_{\tau\in \Omega}\sum_{g_q\in G}C_{g_q}(P_{g_q}^{(\tau)})\label{first_Look-Ahead_General_Ang_Inc}\\
		\mathbf{Subject\;to:}\:\forall(c)\in\mathcal{L},\forall \tau \in \Omega,\;\forall T_r\in \mathbb{T}\notag\\
		P_{{g_q}_{N_i}}^{(\tau)}-P_{{D_d}_{N_i}}^{(\tau)}=\sum_{N_{\overline{i}}\in{J(N_i)}}B_{T_r}^{(c)}(\theta_{N_i}^{(c)(\tau)}-\theta_{N_{\overline{i}}}^{(c)(\tau)});\ \forall{{N_i}\in\mathcal{N}}\label{second I_Look-Ahead_General_Ang_Inc}\\
		|B_{T_r}^{(c)}(\theta_{{T_r}_{t_1}}^{(c)(\tau)}-\theta_{{T_r}_{t_2}}^{(c)(\tau)})|\leq{{\overline{L}}_{T_r}^{(c)}},\;\forall{T_r\in{\mathbb{T}}}\label{fourth_Look-Ahead_General_Ang_Inc}\\
		\underline{R}_{g_q}\leq(P_{g_q}^{(\tau+1)}-P_{g_q}^{(\tau)})\leq \overline{R}_{g_q}\; \forall {g_q}\in G\label{fifth_Look-Ahead_General_Ang_Inc}
		\end{gather}
	\end{subequations}
	In the optimization model (\ref{Look-Ahead_General_Ang_Inc}), the objective function to be minimized, (\ref{first_Look-Ahead_General_Ang_Inc}) represents the total generation cost for all the generators in the system, summed over all the look-ahead dispatch intervals under consideration. The innermost summation is for all the generators and outermost summation is over all the future dispatch intervals. There are three sets of constraints. The first set of constraints (\ref{second I_Look-Ahead_General_Ang_Inc}) describes the nodal power balance. The left-side of the equality represents the nodal power injection, which is the generator output, less the load connected to the particular node and the right-side of the equality represents the power flow along all the transmission lines that are connected to the node (according to the angles represented version of the DC power flow approximation). Inequality constraints (\ref{fourth_Look-Ahead_General_Ang_Inc}) represent the transmission lines' power flow limits for both the base case and each of the contingency scenarios for all the lines and all the dispatch intervals (the base case as well as the scenarios are indexed in the superscript with $(c)$ and the dispatch intervals, by $(\tau)$). As before, the left-side of the inequality represents the DC (or linearized) approximation of the line power flow and the right-side represents the limit. The last set of inequalities (\ref{fifth_Look-Ahead_General_Ang_Inc}) represent the generator ramping constraints, where it can be seen that the change in generator output from one dispatch interval to the other is bounded by the maximum and minimum ramping limits of the the generators.
	\section{APMP Algorithm: The Coarse Grained Decomposition Component}
	\label{APMPAlgo}
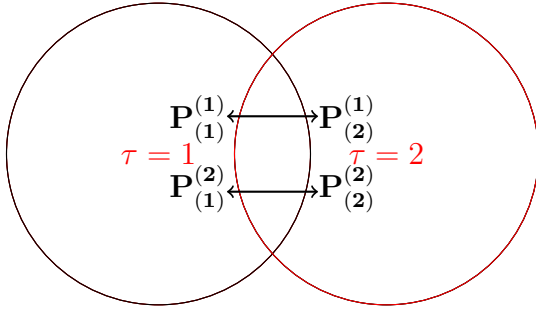
\begin{figure}
		\setlength{\unitlength}{0.4\linewidth}
		\begin{tikzpicture}
		\draw (2,2) circle (2cm) node[anchor=south west] {$\mathbf{P_{(1)}^{(1)}}$};
		\draw[red] (2,2) circle (2cm) node[anchor=center] {$\tau=1$};
		\draw (2,2) circle (2cm) node[anchor=north west] {$\mathbf{P_{(1)}^{(2)}}$};
		\draw (5,2) circle (2cm) node[anchor=north east] {$\mathbf{P_{(2)}^{(2)}}$};
		\draw (5,2) circle (2cm) node[anchor=south east] {$\mathbf{P_{(2)}^{(1)}}$};
		\draw[red] (5,2) circle (2cm) node[anchor=center] {$\tau=2$};
		\draw[thick,<->] (2.9,2.5) -- (4.1,2.5);
		\draw[thick,<->] (2.9,1.5) -- (4.1,1.5);
		\end{tikzpicture}
		\parbox[b]{0.3\linewidth}{
			\caption[Schematic for APP Message Exchange.]
			{Schematic for APP Message Exchange.}
			\label{Schematic_for_APP_Message_Exchange_Appendix}
		}
	\end{figure}

	In this section, we will provide the mathematical formulation of the Auxiliary Proximal Message Passing (APMP) algorithm, as applied to the LASCOPF problem. We will start by describing the APP based coarse-grained component of the algorithm, which will be applied to the LASCOPF problems, followed by the PMP fine-grained component, which decomposes the SCOPF problems (and the OPF problems) of each dispatch interval into the device level computations. The basic idea behind the APP based coarse-grained distributed component of the algorithm is depicted in figure \ref{Schematic_for_APP_Message_Exchange_Appendix}, where the two circles correspond to different dispatch intervals and/or scenarios, across which the computation is split, and the arrows indicate the exchange of messages between those. The messages correspond to the beliefs of each circle or coarse grain about the values of power generations and/or injections/line flows within itself and also of those belonging to the neighboring circles.
	\subsection{Generalized Case of Demand Variation for Multi-Bus Case}
	\label{APP_Demand_Variation_General}
	\begin{figure*}
		\setlength{\unitlength}{0.4\linewidth}
		\begin{tikzpicture}
		\draw[red,thick,dashed] (-1,2) circle (2cm) node[anchor=center] {$\tau=0$};
		\draw[thick,dashed]  (-1,2) circle (2cm) node[anchor=south west] {$\mathbf{P_{g_1}^{(0)}}$};
		\draw[thick,dashed]  (-1,2) circle (2cm) node[anchor=north west] {$\mathbf{P_{g_2}^{(0)}}$};
		\draw (2,2) circle (2cm) node[anchor=south west] {$\mathbf{P_{(1)}^{(1)}}$};
		\draw[red] (2,2) circle (2cm) node[anchor=center] {$\tau=1$};
		\draw (2,2) circle (2cm) node[anchor=north west] {$\mathbf{P_{(1)}^{(2)}}$};
		\draw (5,2) circle (2cm) node[anchor=south west] {$\mathbf{P_{(2)}^{(2)}}$};
		\draw (5,2) circle (2cm) node[anchor=south east] {$\mathbf{P_{(2)}^{(1)}}$};
		\draw[red] (5,2) circle (2cm) node[anchor=center] {$\tau=2$};
		\draw (5,2) circle (2cm) node[anchor=north west] {$\mathbf{P_{(2)}^{(3)}}$};
		\draw (5,2) circle (2cm) node[anchor=north east] {$\mathbf{P_{(2)}^{(2)}}$};
		\draw (8,2) circle (2cm) node[anchor=south east] {$\mathbf{P_{(3)}^{(2)}}$};
		\draw[red] (8,2) circle (2cm) node[anchor=center] {$\tau=3$};
		\draw (8,2) circle (2cm) node[anchor=north east] {$\mathbf{P_{(3)}^{(3)}}$};
		\draw (8,2) circle (2cm) node[anchor=south west] {$\mathbf{P_{g_1}^{(4)}}$};
		\draw (8,2) circle (2cm) node[anchor=north west] {$\mathbf{P_{g_2}^{(4)}}$};
		\draw[thick,dashed] (11,2) circle (2cm) node[anchor=south west] {$\tau=4$};
		\draw[thick,->] (-1,3) -- (1,3);
		\draw[thick,->] (-1,1) -- (1,1);
		\draw[thick,<->] (2.9,2.5) -- (4.1,2.5);
		\draw[thick,<->] (2.9,1.5) -- (4.1,1.5);
		\draw[thick,<->] (5.9,2.5) -- (7.1,2.5);
		\draw[thick,<->] (5.9,1.5) -- (7.1,1.5);
		\draw[thick,->] (9,3) -- (11,3);
		\draw[thick,->] (9,1) -- (11,1);
		\end{tikzpicture}
		\parbox[b]{0.3\linewidth}{
			\caption[Schematic for Look-Ahead SCOPF for Demand Variation]
			{Schematic for Look-Ahead SCOPF for Demand Variation}
			\label{SchematicLookAheadDemandVariation}
		}
	\end{figure*}
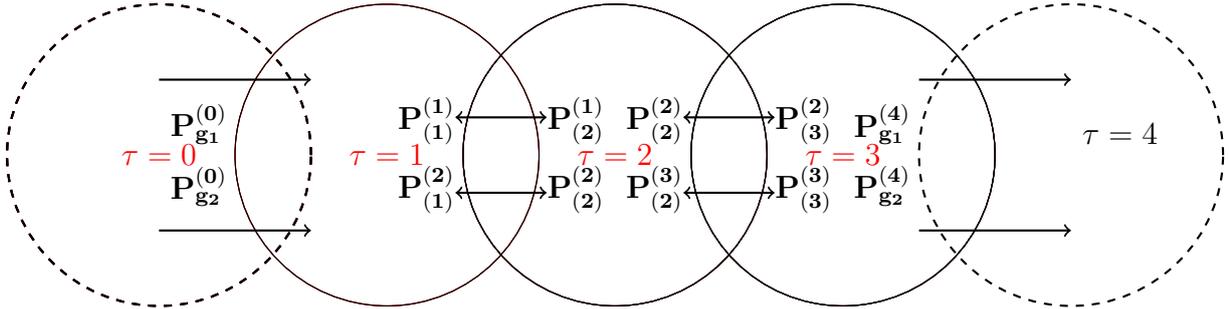
	Referring to the figure~\ref{SchematicLookAheadDemandVariation}, the different overlapping circles here represent the different dispatch intervals, and with $\tau=0$ and $\tau=4$ representing respectively, the latest dispatch interval for which the LASCOPF problem is already solved \& the results are known, and the dispatch interval following the concerned time horizon. In the coarse-grained decomposition, when $\tau = 1$, $P_{g_2(\tau)}^{(\tau-1)}$ and $P_{g_1(\tau)}^{(\tau-1)}$ are replaced by $P_{g_2}^{(0)}$ and $P_{g_1}^{(0)}$ respectively, which are the (known) MW outputs from the current dispatch interval. When $\tau = 3$, $P_{g_2(\tau)}^{(\tau+1)}$ and $P_{g_1(\tau)}^{(\tau+1)}$ are replaced by $P_{g_2}^{(4)}$ and $P_{g_1}^{(4)}$ respectively, which are the last but one iterate values of the MW outputs from the last dispatch interval. We will now apply the Auxiliary Problem Principle (APP)\cite{Cohen:78},  \cite{Cohen:80} to the above optimization problem to derive a set of expressions for the iterate updates. This reformulation is very similar in flavor to the ones presented previously in \cite{KB:97}, \cite{BKCL:99}, \cite{EbrBald:2000}, \cite{BatRen:92} etc. We will state the coarse-grained APP based decomposition of the problem (the one for which the conventional formulation has been presented in equation (\ref{Look-Ahead_General_Ang_Inc})). We will distribute the SCOPF across each dispatch interval and then exchange messages regarding what a particular interval thinks about the optimal values of the decision variables for the intervals immediately preceding and succeeding it, eventually attempting to achieve a consensus between those. In the figure, the values shown at the two sides of the arrows are those ``beliefs" among which we are trying to achieve consensus. We follow the same convention as introduced in section~\ref{notations}. As before, notice that each of the subproblems is very similar to the classical $(N-1)$ SCOPF problems, but with two important differences. First of all, each of these has ramp rate constraints, to account for the change in generator outputs over multiple intervals. Secondly, there are regularization terms added to the cost function for attaining consensus among different coarse grains about the values of the decision variables, which, in this case, are generator outputs. The second terms in the objective functions are the ones representing the proximity from previous iterates. The third and fourth terms in the objective functions of equations (\ref{Look-Ahead_General_First_Interval_Angle}) and (\ref{Look-Ahead_General_Third_Interval_Angle}) and the third, fourth, fifth, and sixth terms in the objective function of equation (\ref{Look-Ahead_General_Second_Interval_Angle}) are the ones for attaining consensus among the coarse grains. The last terms in the objective functions of equations (\ref{Look-Ahead_General_First_Interval_Angle}) and (\ref{Look-Ahead_General_Third_Interval_Angle}) and the last two terms in the objective function of equation (\ref{Look-Ahead_General_Second_Interval_Angle}) correspond to complementary slackness. Note that, in this case also, the real power generation variables are the only ones at which we wish to attain consensus through the application of APP. The bus voltage angles can be thought of as localized to the particular dispatch interval sub-problems, whose values at any iteration are determined by the power generation or injection profiles. We have split the LASCOPF problem into three types of sub-problems, each corresponding to a particular dispatch time interval.\\
	\textbf{First Dispatch Interval SCOPF:}
	\begin{subequations}\label{Look-Ahead_General_First_Interval_Angle}
		\begin{gather}
		\mathbf{P_{(1)}}^{(\mu_{APP}+1)}=\argmin_{\mathbf{P_{(1)}},\theta} \sum_{{g_q}\in G}C_{g_q}(P_{{g_q}(1)}^{(1)})+\notag\\\frac{\beta}{2}||\mathbf{P_{(1)}}-\mathbf{P_{(1)}}^{(\mu_{APP})}||_2^2\notag\\+\gamma[\mathbf{P_{(1)}^{(1)}}^{\dagger}(\mathbf{P_{(1)}^{(1)}}^{(\mu_{APP})}-\mathbf{P_{(2)}^{(1)}}^{(\mu_{APP})})+\notag\\\mathbf{P_{(1)}^{(2)}}^{\dagger}(\mathbf{P_{(1)}^{(2)}}^{(\mu_{APP})}-\mathbf{P_{(2)}^{(2)}}^{(\mu_{APP})})]\notag\\+\mathbf{\lambda_1}^{(\mu_{APP})\dagger}\mathbf{P_{(1)}^{(1)}}+\mathbf{\lambda_2}^{(\mu_{APP})\dagger}\mathbf{P_{(1)}^{(2)}}\label{first_Look-Ahead_General_First_Interval_Angle}\\
		\mathbf{Subject\;to:}\:\forall(c)\in\mathcal{L},\;\forall T_r\in \mathbb{T}\notag\\
		\text{Power-Balance Constraints (Base-Case \& Contingency):}\notag\\
		P_{{g_q}_{N_i}(1)}^{(1)}-P_{{D_d}_{N_i}}^{(1)}=\sum_{N_{\overline{i}}\in{J(N_i)}}B_{T_r}^{(c)}(\theta_{N_i}^{(c)(1)}-\theta_{N_{\overline{i}}}^{(c)(1)});\ \forall{{N_i}\in\mathcal{N}}\label{secondA_Look-Ahead_General_First_Interval_Angle}\\
		\text{Flow Limit Constraints (Base-Case \& Contingency):}\notag\\
		|B_{T_r}^{(c)}(\theta_{{T_r}_{t_1}}^{(c)(1)}-\theta_{{T_r}_{t_2}}^{(c)(1)})|\leq{{\overline{L}}_{T_r}^{(c)}},\;\forall{T_r\in{\mathbb{T}}}\label{third1_Look-Ahead_General_First_Interval_Angle}\\
		\text{Ramp-Rate Constraints:}\notag\\
		\underline{R}_{g_q}\leq P_{{g_q}(1)}^{(2)}-P_{{g_q}(1)}^{(1)}\leq \overline{R}_{g_q},\forall {g_q}\in G\label{fourth_Look-Ahead_General_First_Interval_Angle}\\
		\underline{R}_{g_q}\leq P_{{g_q}(1)}^{(1)}-P_{g_q}^{(0)}\leq \overline{R}_{g_q},\forall {g_q}\in G\label{fifth_Look-Ahead_General_First_Interval_Angle}
		\end{gather}
	\end{subequations}
	In the coarse-grained decomposed sub-problem (\ref{Look-Ahead_General_First_Interval_Angle}), which is for the first look-ahead dispatch interval, the only differences from optimization model (\ref{Look-Ahead_General_Ang_Inc}) are that, in the objective function, (\ref{first_Look-Ahead_General_First_Interval_Angle}) we don't have the summation over the dispatch intervals, for obvious reason, since the computation is split across the dispatch intervals. Also, we can see that, the objective function has some additional regularization terms. And as mentioned in the last paragraph, these are regularization terms added to the cost function for attaining consensus among different coarse grains about the values of the decision variables, which, in this case, are generator outputs. The second terms in the objective function is the one representing the proximity from previous iterate, $\mathbf{P_{(1)}}^{(\mu_{APP})}$. The third and fourth terms in the objective function (\ref{first_Look-Ahead_General_First_Interval_Angle}) are the ones for attaining consensus among the coarse grains. That is, in the third term, attaining consensus between $\mathbf{P_{(1)}^{(1)}}^{(\mu_{APP})}$, which are the beliefs of interval 1 about its own generator outputs and $\mathbf{P_{(2)}^{(1)}}^{(\mu_{APP})}$, which are the beliefs of interval 2 about the generator outputs of interval 1.  The fourth term has a similar interpretation as to attaining consensus among interval 1's beliefs of generator outputs of interval 2 and interval 2's beliefs about its own. The last two terms in the objective function correspond to complementary slackness. Note that, in this case also, the real power generation variables are the only ones at which we wish to attain consensus through the application of APP. The bus voltage angles can be thought of as localized to the particular dispatch interval sub-problems, whose values at any iteration are determined by the power generation or injection profiles. The rest of the constraints corresponding to the nodal power balance, line flow limits, and generator ramping are exactly the same as those appearing in (\ref{Look-Ahead_General_Ang_Inc}).\\
	\textbf{Intermediate Dispatch Interval $\tau \in \{2,3,...,|\Omega|-1\}$ SCOPF:}
	\begin{subequations}\label{Look-Ahead_General_Second_Interval_Angle}
		\begin{gather}
		\mathbf{P_{(\tau)}}^{(\mu_{APP}+1)}=\argmin_{\mathbf{P_{(\tau)}},\theta} \sum_{{g_q}\in G}C_{g_q}(P_{{g_q}(\tau)}^{(\tau)})+\notag\\\frac{\beta}{2}||\mathbf{P_{(\tau)}}-\mathbf{P_{(\tau)}}^{(\mu_{APP})}||_2^2\notag\\+\gamma[\mathbf{P_{(\tau)}^{(\tau-1)}}^{\dagger}(\mathbf{P_{(\tau)}^{(\tau-1)}}^{(\mu_{APP})}-\mathbf{P_{(\tau-1)}^{(\tau-1)}}^{(\mu_{APP})})+\notag\\\mathbf{P_{(\tau)}^{(\tau)}}^{\dagger}(\mathbf{P_{(\tau)}^{(\tau)}}^{(\mu_{APP})}-\mathbf{P_{(\tau-1)}^{(\tau)}}^{(\mu_{APP})})\notag\\+\mathbf{P_{(\tau)}^{(\tau)}}^{\dagger}(\mathbf{P_{(\tau)}^{(\tau)}}^{(\mu_{APP})}-\mathbf{P_{(\tau+1)}^{(\tau)}}^{(\mu_{APP})})+\notag\\\mathbf{P_{(\tau)}^{(\tau+1)}}^{\dagger}(\mathbf{P_{(\tau)}^{(\tau+1)}}^{(\mu_{APP})}-\mathbf{P_{(\tau+1)}^{(\tau+1)}}^{(\mu_{APP})})]\notag\\+ \mathbf{\lambda_{2\tau-1}}^{(\mu_{APP})\dagger}\mathbf{P_{(\tau)}^{(\tau)}}+\mathbf{\lambda_{2\tau}}^{(\mu_{APP})\dagger}\mathbf{P_{(\tau)}^{(\tau+1)}}\notag\\-\mathbf{\lambda_{2\tau-3}}^{(\mu_{APP})\dagger}\mathbf{P_{(\tau)}^{(\tau-1)}}-\mathbf{\lambda_{2\tau-2}}^{(\mu_{APP})\dagger}\mathbf{P_{(\tau)}^{(\tau)}}\label{first_Look-Ahead_General_Second_Interval_Angle}\\
		\mathbf{Subject\;to:}\:\forall(c)\in\mathcal{L},\;\forall T_r\in \mathbb{T}\notag\\
		\text{Power-Balance Constraints (Base-Case \& Contingency):}\notag\\
		P_{{g_q}_{N_i}(\tau)}^{(\tau)}-P_{{D_d}_{N_i}}^{(\tau)}=\sum_{N_{\overline{i}}\in{J(N_i)}}B_{T_r}^{(c)}(\theta_{N_i}^{(c)(\tau)}-\theta_{N_{\overline{i}}}^{(c)(\tau)});\ \forall{{N_i}\in\mathcal{N}}\label{secondA_Look-Ahead_General_Second_Interval_Angle}\\
		\text{Flow Limit Constraints (Base-Case \& Contingency):}\notag\\
		|B_{T_r}^{(c)}(\theta_{{T_r}_{t_1}}^{(c)(\tau)}-\theta_{{T_r}_{t_2}}^{(c)(\tau)})|\leq{{\overline{L}}_{T_r}^{(c)}},\;\forall{T_r\in{\mathbb{T}}}\label{third1_Look-Ahead_General_Second_Interval_Angle}\\
		\text{Ramp-Rate Constraints:}\notag\\
		\underline{R}_{g_q}\leq P_{{g_q}(\tau)}^{(\tau+1)}-P_{{g_q}(\tau)}^{(\tau)}\leq \overline{R}_{g_q},\forall {g_q}\in G\label{fourth_Look-Ahead_General_Second_Interval_Angle}\\
		\underline{R}_{g_q}\leq P_{{g_q}(\tau)}^{(\tau)}-P_{{g_q}(\tau)}^{(\tau-1)}\leq \overline{R}_{g_q},\forall {g_q}\in G\label{fifth_Look-Ahead_General_Second_Interval_Angle}
		\end{gather}
	\end{subequations}
	In the coarse-grained decomposed sub-problem (\ref{Look-Ahead_General_Second_Interval_Angle}), which are for the intermediate dispatch intervals, as opposed to the objective function for the optimization problem for the first dispatch interval, in  (\ref{first_Look-Ahead_General_Second_Interval_Angle}) there are four regularization terms after the one representing the proximity from previous iterate, $\mathbf{P_{(\tau)}}^{(\mu_{APP})}$. The third term is for attaining consensus between $\mathbf{P_{(\tau)}^{(\tau-1)}}^{(\mu_{APP})}$, which are the beliefs of interval $\tau$ about the generator outputs of interval $\tau-1$ and $\mathbf{P_{(\tau-1)}^{(\tau-1)}}^{(\mu_{APP})}$, which are the beliefs of interval $\tau-1$ about its own generator outputs.  The fourth term has a similar interpretation as to attaining consensus among interval $\tau$'s beliefs of its own generator outputs and interval $(\tau-1)$'s beliefs about generator outputs of interval $\tau$, whereas the fifth term pertains to attaining consensus among interval $\tau$'s beliefs of its own generator outputs and interval $(\tau+1)$'s beliefs about generator outputs of interval $\tau$. The sixth term pertains to attaining consensus among the belief of interval $\tau$ about generator outputs of interval $\tau+1$ and belief of interval $\tau+1$ about its own generator outputs. The last four terms in the objective function correspond to complementary slackness. The rest of the constraints corresponding to the nodal power balance, line flow limits, and generator ramping are again, exactly the same as those appearing in (\ref{Look-Ahead_General_Ang_Inc}).\\
	\textbf{Last Dispatch Interval, or Dispatch Interval $|\Omega|$ SCOPF:}
	\begin{subequations}\label{Look-Ahead_General_Third_Interval_Angle}
		\begin{gather}
		\mathbf{P_{(|\Omega|)}}^{(\mu_{APP}+1)}= \argmin_{\mathbf{P_{(|\Omega|)}},\theta} \sum_{{g_q}\in G}C_{g_q}(P_{{g_q}(|\Omega|)}^{(|\Omega|)})+\notag\\ \frac{\beta}{2}||\mathbf{P_{(|\Omega|)}}-\mathbf{P_{(|\Omega|)}}^{(\mu_{APP})}||_2^2\notag\\+ \gamma[\mathbf{P_{(|\Omega|)}^{(|\Omega|-1)}}^{\dagger}(\mathbf{P_{(|\Omega|)}^{(|\Omega|-1)}}^{(\mu_{APP})}-\mathbf{P_{(|\Omega|-1)}^{(|\Omega|-1)}}^{(\mu_{APP})})+\notag\\\mathbf{P_{(|\Omega|)}^{(|\Omega|)}}^{\dagger}(\mathbf{P_{(|\Omega|)}^{(|\Omega|)}}^{(\mu_{APP})}-\mathbf{P_{(|\Omega|-1)}^{(|\Omega|)}}^{(\mu_{APP})})]\notag\\- \mathbf{\lambda_{2|\Omega|-3}}^{(\mu_{APP})\dagger}\mathbf{P_{(|\Omega|)}^{(|\Omega|-1)}}-\mathbf{\lambda_{2|\Omega|-2}}^{(\mu_{APP})\dagger}\mathbf{P_{(|\Omega|)}^{(|\Omega|)}}\label{first_Look-Ahead_General_Third_Interval_Angle}\\
		\mathbf{Subject\;to:}\:\forall(c)\in\mathcal{L},\;\forall T_r\in \mathbb{T}\notag\\
		\text{Power-Balance Constraints (Base-Case \& Contingency):}\notag\\
		P_{{g_q}_{N_i}(|\Omega|)}^{(|\Omega|)}-P_{{D_d}_{N_i}}^{(|\Omega|)}=\sum_{N_{\overline{i}}\in{J(N_i)}}B_{T_r}^{(c)}(\theta_{N_i}^{(c)(|\Omega|)}-\theta_{N_{\overline{i}}}^{(c)(|\Omega|)});\notag\\\forall{{N_i}\in\mathcal{N}}\label{secondA_Look-Ahead_General_Third_Interval_Angle}\\
		\text{Flow Limit Constraints (Base-Case \& Contingency):}\notag\\
		|B_{T_r}^{(c)}(\theta_{{T_r}_{t_1}}^{(c)(|\Omega|)}-\theta_{{T_r}_{t_2}}^{(c)(|\Omega|)})|\leq{{\overline{L}}_{T_r}^{(c)}},\;\forall{T_r\in{\mathbb{T}}}\label{third1_Look-Ahead_General_Third_Interval_Angle}\\
		\text{Ramp-Rate Constraints:}\notag\\
		\underline{R}_{g_q}\leq P_{{g_q}(|\Omega|)}^{(|\Omega|)(\mu_{APP})}-P_{{g_q}(|\Omega|)}^{(|\Omega|)}\leq \overline{R}_{g_q},\forall {g_q}\in G\label{fourth_Look-Ahead_General_Third_Interval_Angle}\\
		\underline{R}_{g_q}\leq P_{{g_q}(|\Omega|)}^{(|\Omega|)}-P_{{g_q}(|\Omega|)}^{(|\Omega|-1)}\leq \overline{R}_{g_q},\forall {g_q}\in G\label{fifth_Look-Ahead_General_Third_Interval_Angle}\\
		\mathbf{Dual\:Variable \:Updates:}\notag\\
		\mathbf{\lambda_{\tau}}^{(\mu_{APP}+1)}=\mathbf{\lambda_{\tau}}^{(\mu_{APP})}+\alpha(\mathbf{P_{({\tau})}^{({\tau})}}^{(\mu_{APP}+1)}-\mathbf{P_{({\tau}+1)}^{({\tau})}}^{(\mu_{APP}+1)})\label{eighth_Look-Ahead_Simple_APP}\\
		\mathbf{\lambda_{\tau+1}}^{(\mu_{APP}+1)}=\mathbf{\lambda_{\tau+1}}^{(\mu_{APP})}+\alpha(\mathbf{P_{({\tau})}^{({\tau+1})}}^{(\mu_{APP}+1)}-\mathbf{P_{({\tau+1})}^{({\tau+1})}}^{(\mu_{APP}+1)})\label{ninth_Look-Ahead_Simple_APP}
		\end{gather}
	\end{subequations}
	Lastly, in the coarse-grained decomposed sub-problem (\ref{Look-Ahead_General_Third_Interval_Angle}), which is for the last dispatch interval under consideration, in the objective function for the optimization problem (\ref{first_Look-Ahead_General_Third_Interval_Angle}) the two regularization terms after the one representing the proximity from previous iterate, $\mathbf{P_{(|\Omega|)}}^{(\mu_{APP})}$. The third term is for attaining consensus between $\mathbf{P_{(|\Omega|)}^{(|\Omega|-1)}}^{(\mu_{APP})}$, which are the beliefs of interval $|\Omega|$ about the generator outputs of interval $|\Omega|-1$ and $\mathbf{P_{(|\Omega|-1)}^{(|\Omega|-1)}}^{(\mu_{APP})}$, which are the beliefs of interval $|\Omega|-1$ about its own generator outputs.  The fourth term has a similar interpretation as to attaining consensus among interval $|\Omega|$'s beliefs of its own generator outputs and interval $(|\Omega|-1)$'s beliefs about generator outputs of interval $|\Omega|$. The last two terms in the objective function correspond to complementary slackness. The rest of the constraints corresponding to the nodal power balance, line flow limits, and generator ramping are again, exactly the same as those appearing in (\ref{Look-Ahead_General_Ang_Inc}). Equations (\ref{eighth_Look-Ahead_Simple_APP}) and (\ref{ninth_Look-Ahead_Simple_APP}) pertain to updates of the dual variables or the Lagrange multipliers, $\lambda$s corresponding to the different power generation consensus terms described above. The lack of consensus gives the step direction, while $\alpha$ is the step length, as can be seen above.\\
	We have presented a coarse grained parallelization of the look-ahead SCOPF to cope with demand variation, where the entire problem is decomposed across the different dispatch time intervals. Subsequently, we will present the ADMM-Proximal Message Passing based fine grained decomposition and parallelization of the problem.
	\section{$\mathcal{DTN}$ Reformulations of the SCOPF Problems}
	\label{DTNFormulation}
	In this section we carry out the reformulation of the SCOPF models that we presented in the last section within each APP coarse grain, in order for us to be able to solve those problems by the Proximal Message Passing method. In order to simplify the presentation and also better clarify the meaning of each term, we will group the terms of the objective into four different categories. We will define them and write down the expressions pertaining to the SCOPF in the coarse grain corresponding to equation (\ref{Look-Ahead_General_Second_Interval_Angle}). These are:\\
	\textbf{1) Cost of Generation} ($C(P)$): This term consists of the actual total cost of generating real power by the different generators as well as the indicator functions corresponding to the lower and upper generating limits of the different generators. For this term, the real power generated is always considered at the base case. This term is given as:\\
	$C(\mathbf{P^{(0)(\tau)}})=\sum_{t_k\in g_q\cap{\mathcal{T}}, q=1}^{|G|}\bigg(C_{g_q}({P}_{{g_q}_{t_k}(\tau)}^{(0)(\tau)})+I_{\leq}(\overline{P}_{g_q}-{P}_{{g_q}_{t_k}(\tau)}^{(0)(\tau)})
	+I_{\leq}({P}_{{g_q}_{t_k}(\tau)}^{(0)(\tau)}-\underline{P}_{g_q})+\frac{\beta}{2}\bigg[({P}_{{g_q}_{t_k}(\tau)}^{(0)(\tau)}-{P}_{{g_q}_{t_k}(\tau)}^{(0)(\tau)(\mu_{APP})})^2+\sum_{s=-1,1}({P}_{{g_q}_{t_k}(\tau)}^{(0)(\tau+s)}-{P}_{{g_q}_{t_k}(\tau)}^{(0)(\tau+s)(\mu_{APP})})^2\bigg]+\gamma\bigg[\sum_{s=-1,1}({P}_{{g_q}_{t_k}(\tau)}^{(0)(\tau)}({P}_{{g_q}_{t_k}(\tau)}^{(0)(\tau)(\mu_{APP})}-{P}_{{g_q}_{t_k}(0)(\tau+s)}^{(0)(\tau)(\mu_{APP})})+{P}_{{g_q}_{t_k}(\tau)}^{(0)(\tau+s)}({P}_{{g_q}_{t_k}(\tau)}^{(0)(\tau+s)(\mu_{APP})}-{P}_{{g_q}_{t_k}(0)(\tau+s)}^{(0)(\tau+s)(\mu_{APP})}))\bigg]+(\lambda_{{g_q}2(\tau-1)}^{(\mu_{APP})}-\lambda_{{g_q}(2\tau-2)}^{(\mu_{APP})}){P}_{{g_q}_{t_k}(\tau)}^{(0)(\tau)}+\lambda_{{g_q}(2\tau)}^{(\mu_{APP})}{P}_{{g_q}_{t_k}(\tau)}^{(0)(\tau+1)}-\lambda_{{g_q}(2\tau-3)}^{(\mu_{APP})}{P}_{{g_q}_{t_k}(\tau)}^{(0)(\tau-1)}\bigg)$\\
	For the sake of brevity, we are presenting here the terms corresponding to line flow limit constraint ($F(P^{(c)(\tau)})$), power-angle relation ($\chi(P^{(c)(\tau)},\theta^{(c)(\tau)})$), and ramp constraint ($\Delta(P^{(0)(\tau)})$) in the most general and condensed forms. The details of these can be found in \cite{CK:14} and \cite{Sambuddha2017}.\\
	\textbf{2) Line Flow Limit Constraint} ($F(P)$): This term consists of the sum of the indicator functions corresponding to enforcing that real power flows on the lines are less than the maximum allowed, both at the base-case as well as under different contingencies. This term can be written as:\\
	$F(P^{(c)(\tau)})=\sum_{(c)\in\mathcal{L}}\sum_{T_r\in \mathbb{T}}\sum_{t_k\in T_r\cap{\mathcal{T}}}I_{\leq}({\overline{L}}_{T_r}^{(c)}-|{P}_{{T_r}_{t_k}}^{(c)(\tau)}|)$\\
	\textbf{3) Power-Angle Relation} ($\chi(P,\theta)$):This term consists of the sum of the indicator functions corresponding to the relation of the power flow at each end of the lines and the voltage phase angles at the two ends, both at the base-case and the contingencies. This term can be written as:\\
	$\chi(P^{(c)(\tau)},\theta^{(c)(\tau)})=\sum_{(c)\in\mathcal{L}}\sum_{T_r\in \mathbb{T}}\sum_{t_k,t_{k^{'}}\in T_r\cap{\mathcal{T}}}I_{=}\Bigg({P}_{{T_r}_{t_k}}^{(c)(\tau)}+\frac{{\theta}_{{T_r}_{t_k}}^{(c)(\tau)}-{\theta}_{{T_r}_{t_{k^{'}}}}^{(c)(\tau)}}{{X}_{T_r}^{(c)}}\Bigg)$\\
	\textbf{4) Ramp Constraint} ($\Delta(P)$):This term corresponds to the change of power output of generator from one time period to another and the maximum rate at which it can go up or down. This term can be written as:\\
	$\Delta(P^{(0)(\tau)})=\sum_{s=-1,1}\sum_{t_k\in g_q\cap{\mathcal{T}}, q=1}^{|G|}(I_{\leq}(\overline{R}_{g_q}-P_{{g_q}_{t_k}(\tau)}^{(0)(\tau+s)}+P_{{g_q}_{t_k}(\tau)}^{(0)(\tau)})+I_{\leq}(P_{{g_q}_{t_k}(\tau)}^{(0)(\tau+s)}-P_{{g_q}_{t_k}(\tau)}^{(0)(\tau)}-\underline{R}_{g_q}))$\\
	The reformulated equations for the problem stated in equations (\ref{Look-Ahead_General_First_Interval_Angle})-(\ref{Look-Ahead_General_Third_Interval_Angle}) are as follows:
	\begin{subequations}\label{DTN_for_Contingency_General_Looakahead_Demand}
		\begin{gather}
		\min_{P_{t_k}^{(c)(\tau)}, \theta_{t_k}^{(c)(\tau)}}f(P)=C(P^{(0)(\tau)})
		+F(P^{(c)(\tau)})+\notag\\\chi(P^{(c)(\tau)},\theta^{(c)(\tau)})
		+\Delta(P^{(0)(\tau)})\label{DTN_for_Contingency_General_Looakahead_Demand1}\\
		\text{Subject to: }\hat{P}_{{N_i}_{t_k}}^{(c)(\tau)}=0, \tilde{\theta}_{{N_i}_{t_k}}^{(c)(\tau)}=0,\notag\\\forall N_i\in\mathcal{N}, \forall t_k\in\mathcal{T}, \forall (c)\in\mathcal{L}\label{DTN_for_Contingency_General_Looakahead_Demand2}
		\end{gather}
	\end{subequations}
	In the above set of equations, $\hat{P}_{{N_i}_{t_k}}^{(c)(\tau)}$ is the average node power injection and $\tilde{\theta}_{{N_i}_{t_k}}^{(c)(\tau)}$ is the deviation of the voltage phase angle of each device connected to a particular node from the average node voltage phase angle. Readers are referred to \cite{CK:14}, \cite{KC:13} for more information on these. Hence, constraints stated in equation (\ref{DTN_for_Contingency_General_Looakahead_Demand2}) represent the Kirchhoff's law for node power balance and angle consistency.
	\section{APMP Algorithm: The Fine Grained Decomposition Component}
	\label{PMPAlgorithm}
	In this section we state the equations for the Proximal Message Passing algorithm based fine-grained decomposition of the above problem.
	\subsection{Proximal Message Passing for the Look-Ahead Dispatch: Generalized Case of Demand Variation for the Multi Bus-Case}
	From equation (\ref{DTN_for_Contingency_General_Looakahead_Demand}), the slightly reformulated $\mathcal{DTN}$ equations are as follows:
	\begin{equation}\label{DTN_Reform_Demand_Var_General}
	\begin{array}{ll}
	\min_{P_{t_k}^{(c)},\theta_{t_k}^{(c)}}C(P^{(0)(\tau)})
	+F(P^{(c)(\tau)})+\chi(P^{(c)(\tau)},\theta^{(c)(\tau)})+\\\Delta(P^{(0)(\tau)})+\sum_{(c)\in\mathcal{L}}\sum_{{N_i}\in\mathcal{N}}(\hat{I}(z_{{N_i}{t_k}}^{(c)(\tau)})+\tilde{I}(\xi_{{N_i}{t_k}}^{(c)(\tau)}))\\
	\text{Subject to: }P_{t_k}^{(c)(\tau)}=z_{t_k}^{(c)(\tau)}, {\theta}_{t_k}^{(c)(\tau)}=\xi_{t_k}^{(c)(\tau)},\\ \forall N_i\in\mathcal{N}, \forall t_k\in\mathcal{T},
	\forall (c)\in\mathcal{L}
	\end{array}
	\end{equation}
	where $\hat{I}(z_{{N_i}{t_k}})$ and $\tilde{I}(\xi_{{N_i}{t_k}})$ are indicator functions of the sets $\{z_{t_k}|\hat{z}_{{N_i}{t_k}}=0\}$ and $\{\xi_{t_k}|\tilde{\xi}_{{N_i}{t_k}}=0\}$ respectively. Again, for the sake of brevity, we will skip writing down the iterates for the transmission lines and loads (which are simple projections), and make a mention that they are the same as the ones presented in references \cite{CK:14} and \cite{Sambuddha2017}.
	\subsubsection{Iterates for Generators}The iterates for generators consist of the update equations for the real power output and voltage-phase angles of the generator terminals for both the base case and the different $(N-1)$ contingency scenarios and are as follows:
	\begin{equation}\label{ADMM_Lookahead_Demand_Gen}
	\begin{array}{ll}
	(P_{{g_q}{t_k}}^{(0)(\tau)(\nu+1)},\theta_{{g_q}{t_k}}^{(c)(\tau)(\nu+1)})\\=\mbox{argmin}_{P_{{g_q}{t_k}}^{(0)(\tau)},\theta_{{g_q}{t_k}}^{(c)(\tau)}}[C_{g_q}({P}_{{g_q}_{t_k}(\tau)}^{(0)(\tau)})+I_{\leq}(\overline{P}_{g_q}-{P}_{{g_q}_{t_k}(\tau)}^{(0)(\tau)})
	\notag\\+I_{\leq}({P}_{{g_q}_{t_k}(\tau)}^{(0)(\tau)}-\underline{P}_{g_q})+\frac{\beta}{2}[({P}_{{g_q}_{t_k}(\tau)}^{(0)(\tau)}-{P}_{{g_q}_{t_k}(\tau)}^{(0)(\tau)(\mu_{APP})})^2\notag\\+\sum_{s=-1,1}({P}_{{g_q}_{t_k}(\tau)}^{(0)(\tau+s)}-{P}_{{g_q}_{t_k}(\tau)}^{(0)(\tau+s)(\mu_{APP})})^2]+\notag\\\gamma[\sum_{s=-1,1}({P}_{{g_q}_{t_k}(\tau)}^{(0)(\tau)}({P}_{{g_q}_{t_k}(\tau)}^{(0)(\tau)(\mu_{APP})}-{P}_{{g_q}_{t_k}(0)(\tau+s)}^{(0)(\tau)(\mu_{APP})})\notag\\+{P}_{{g_q}_{t_k}(\tau)}^{(0)(\tau+s)}({P}_{{g_q}_{t_k}(\tau)}^{(0)(\tau+s)(\mu_{APP})}-{P}_{{g_q}_{t_k}(0)(\tau+s)}^{(0)(\tau+s)(\mu_{APP})}))]+\notag\\(\lambda_{{g_q}2(\tau-1)}^{(\mu_{APP})}-\lambda_{{g_q}(2\tau-2)}^{(\mu_{APP})}){P}_{{g_q}_{t_k}(\tau)}^{(0)(\tau)}+\lambda_{{g_q}(2\tau)}^{(\mu_{APP})}{P}_{{g_q}_{t_k}(\tau)}^{(0)(\tau+1)}-\notag\\\lambda_{{g_q}(2\tau-3)}^{(\mu_{APP})}{P}_{{g_q}_{t_k}(\tau)}^{(0)(\tau-1)}+\Delta(P^{(0)(\tau)})+\notag\\
	\sum_{(c)\in\mathcal{L}}(\frac{\rho}{2})({||P_{{g_q}_{t_k}}^{(0)(\tau)}-z_{{g_q}{t_k}}^{(c)(\tau)(\nu)}+u_{{g_q}{t_k}}^{(c)(\tau)(\nu)}||}_{2}^{2}\\+{||\theta_{{g_q}_{t_k}}^{(c)(\tau)}-\xi_{{g_q}{t_k}}^{(c)(\tau)(\nu)}+v_{{g_q}{t_k}}^{(c)(\tau)(\nu)}||}_{2}^{2})],\\
	\forall g_q\in G, \tau\in\Omega,t_k\in\mathcal{T}\cap G
	\end{array}
	\end{equation}
	\subsubsection{Iterates for Nets}
\begin{figure}
		\begin{center}
			\vspace*{-4cm}
			\includegraphics[height=15cm,width=20cm]{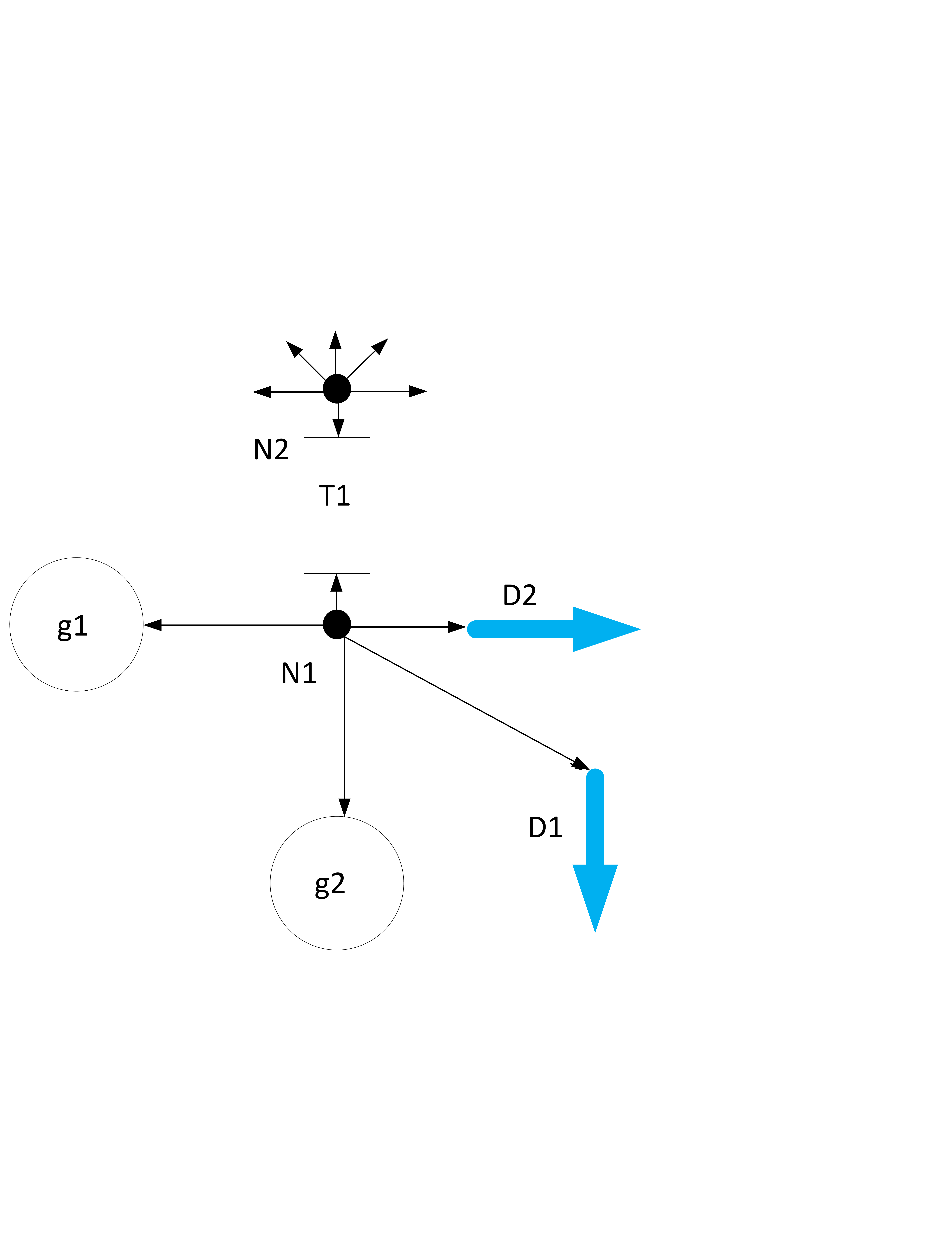}
			\caption{Broadcast: Calculation of Prox Function for Devices}
			\label{Broadcast}
		\end{center}
	\end{figure}

\begin{figure}
		\begin{center}
			\vspace*{-4cm}
			\includegraphics[height=15cm,width=20cm]{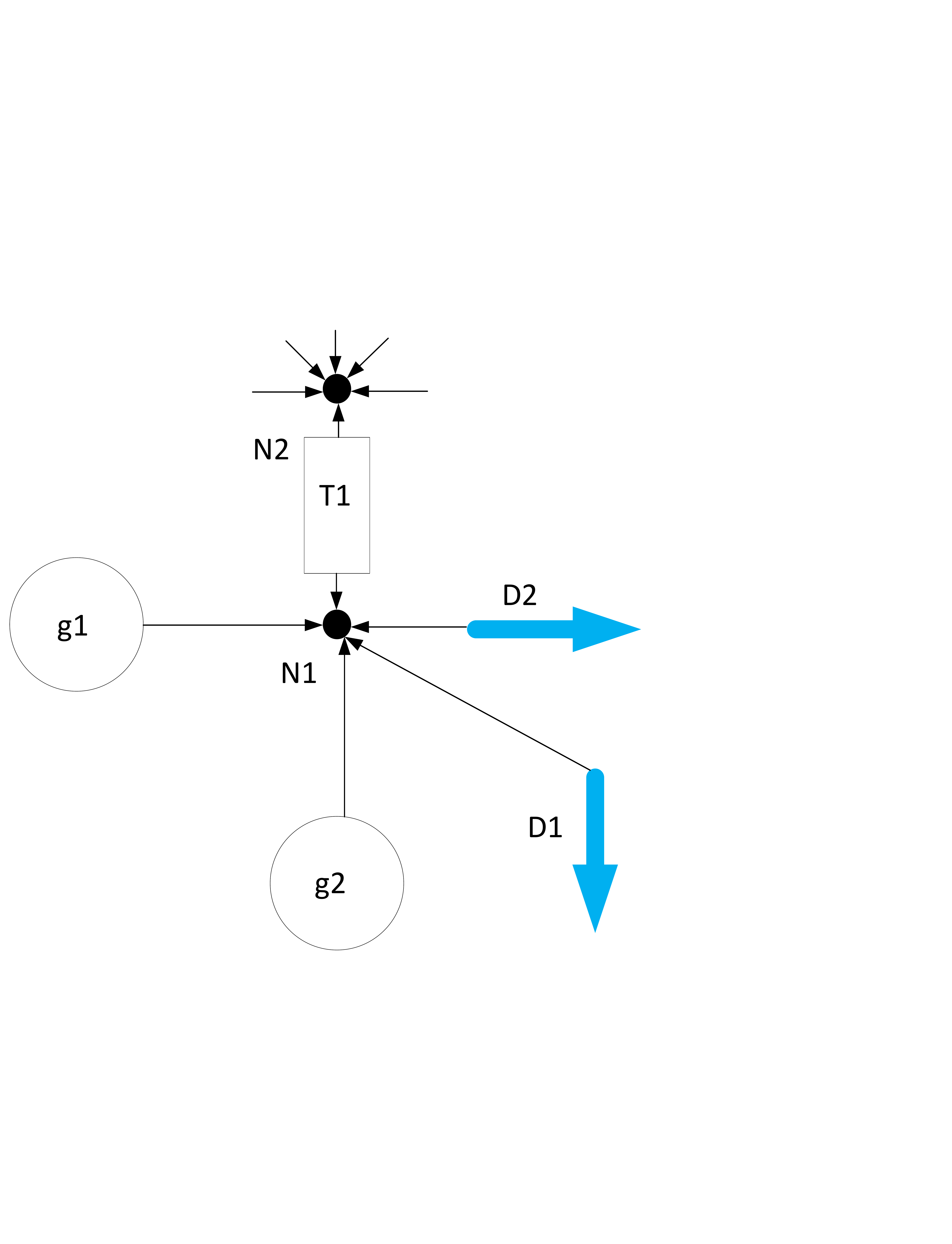}
			\caption{Gather: Updating of Dual variables by Nodes}
			\label{Gather}
		\end{center}
	\end{figure}

\begin{figure}
		\begin{center}
			\vspace*{-2cm}
			\hspace*{-4cm}
			\includegraphics[height=15cm,width=20cm]{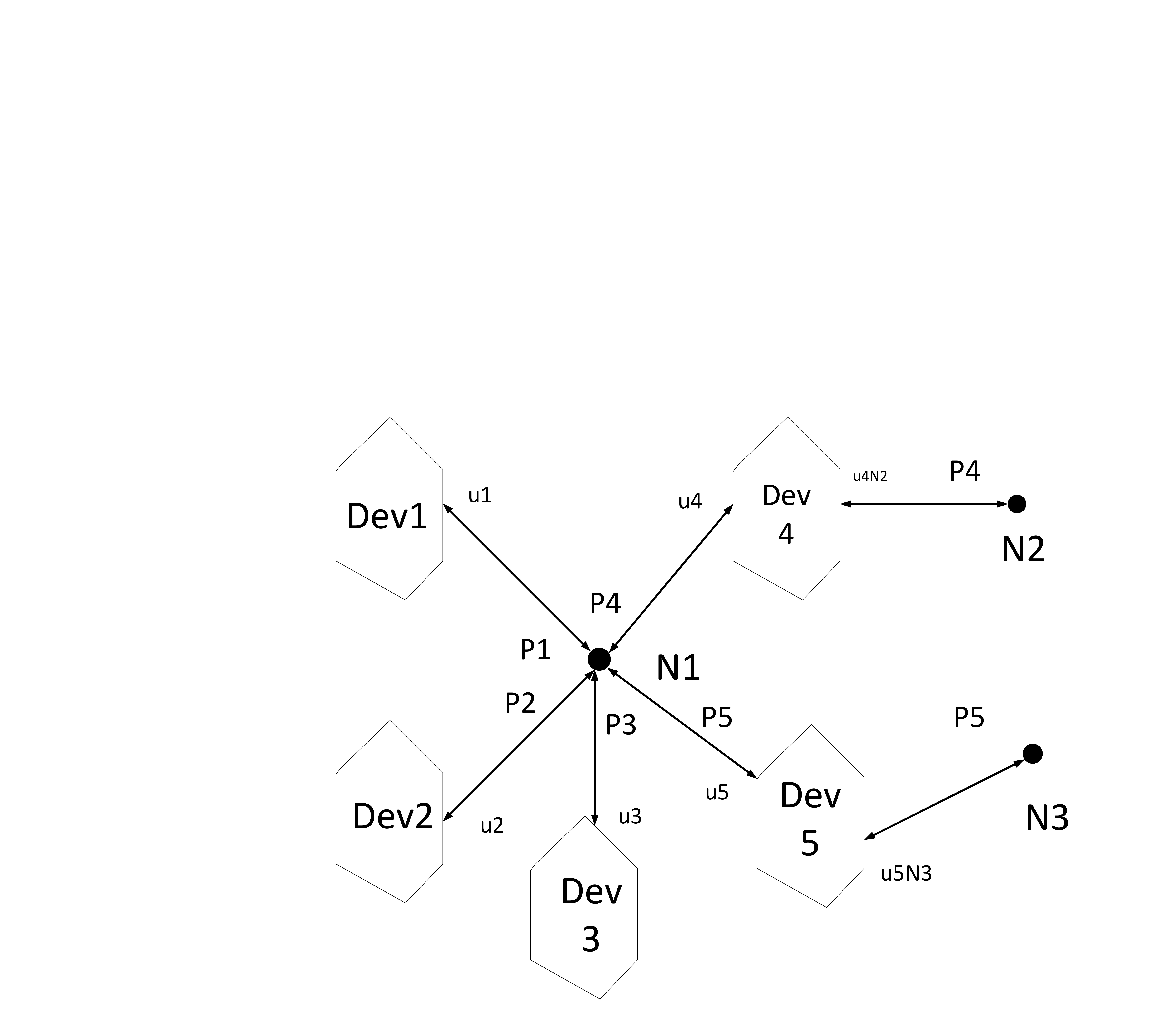}
			\caption{Generic ADMM: Broadcast and Gather}
			\label{ADMMGen}
		\end{center}
	\end{figure}

\begin{figure}
		\begin{center}
			\vspace*{-2cm}
			\hspace*{-4cm}
			\includegraphics[width=0.92\linewidth,trim=5mm 12mm 5mm 5mm, clip]{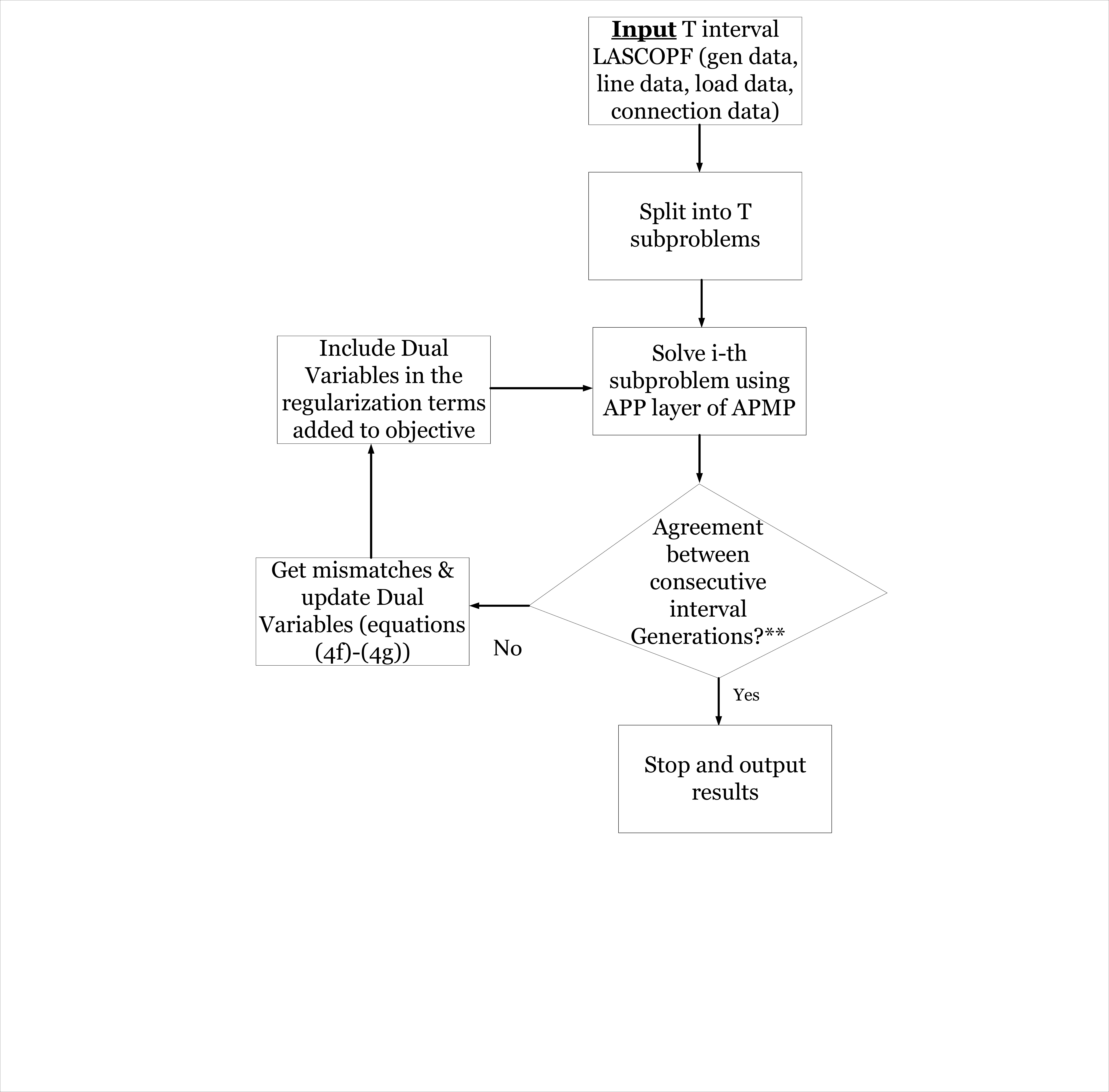}
			\caption{Flowchart for the outermost APP layer of APMP for solving LASCOPF}
			\label{outermostAPP}
		\end{center}
	\end{figure}

\begin{figure}
		\begin{center}
			\vspace*{-2cm}
			\hspace*{-4cm}
			\includegraphics[width=0.92\linewidth,trim=5mm 12mm 5mm 5mm, clip]{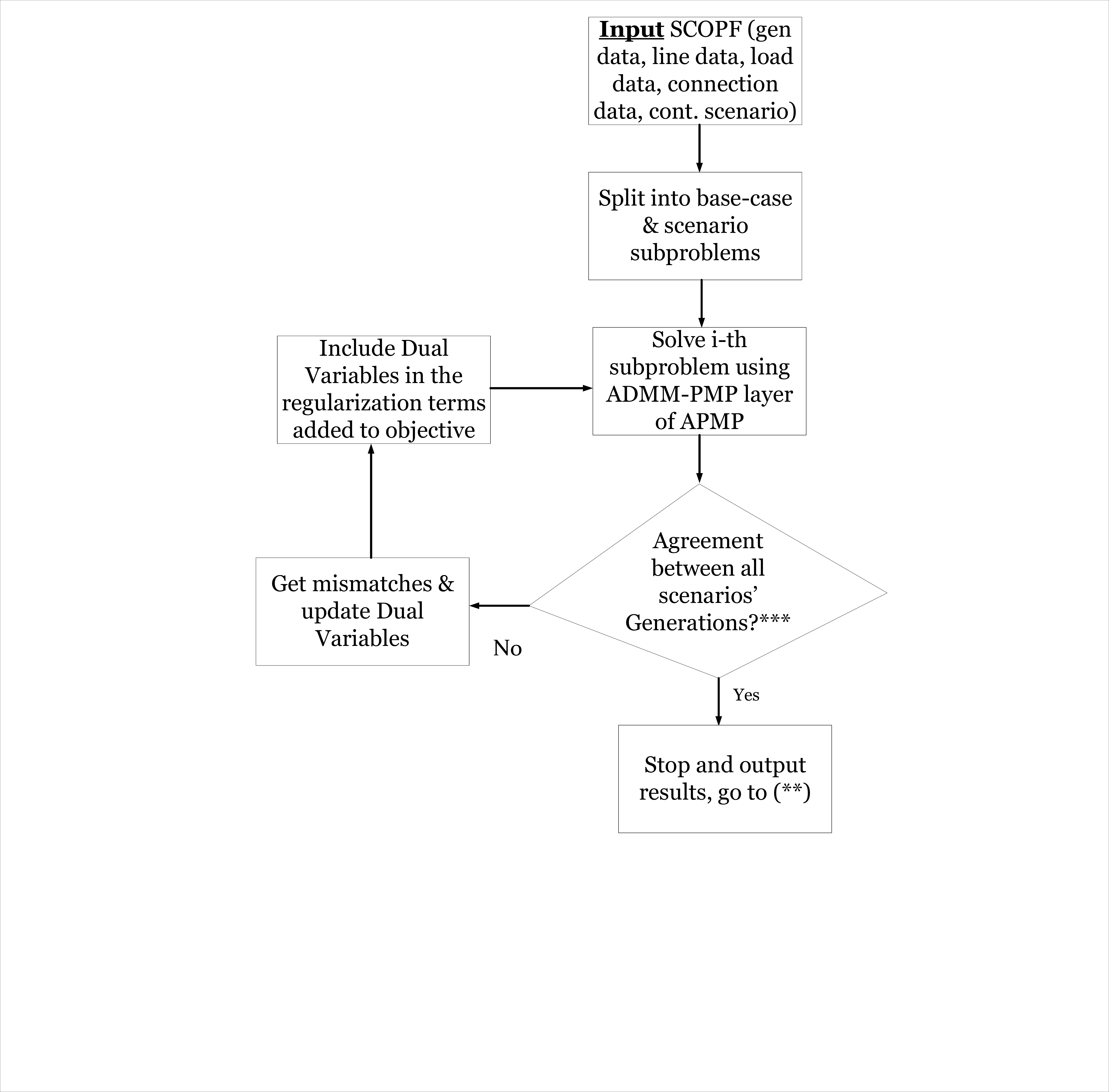}
			\caption{Flowchart for the inner APP layer for solving SCOPF}
			\label{SCOPFAPP}
		\end{center}
	\end{figure}

\begin{figure}
		\begin{center}
			\vspace*{-2cm}
			\hspace*{-4cm}
			\includegraphics[width=0.92\linewidth,trim=5mm 12mm 5mm 5mm, clip]{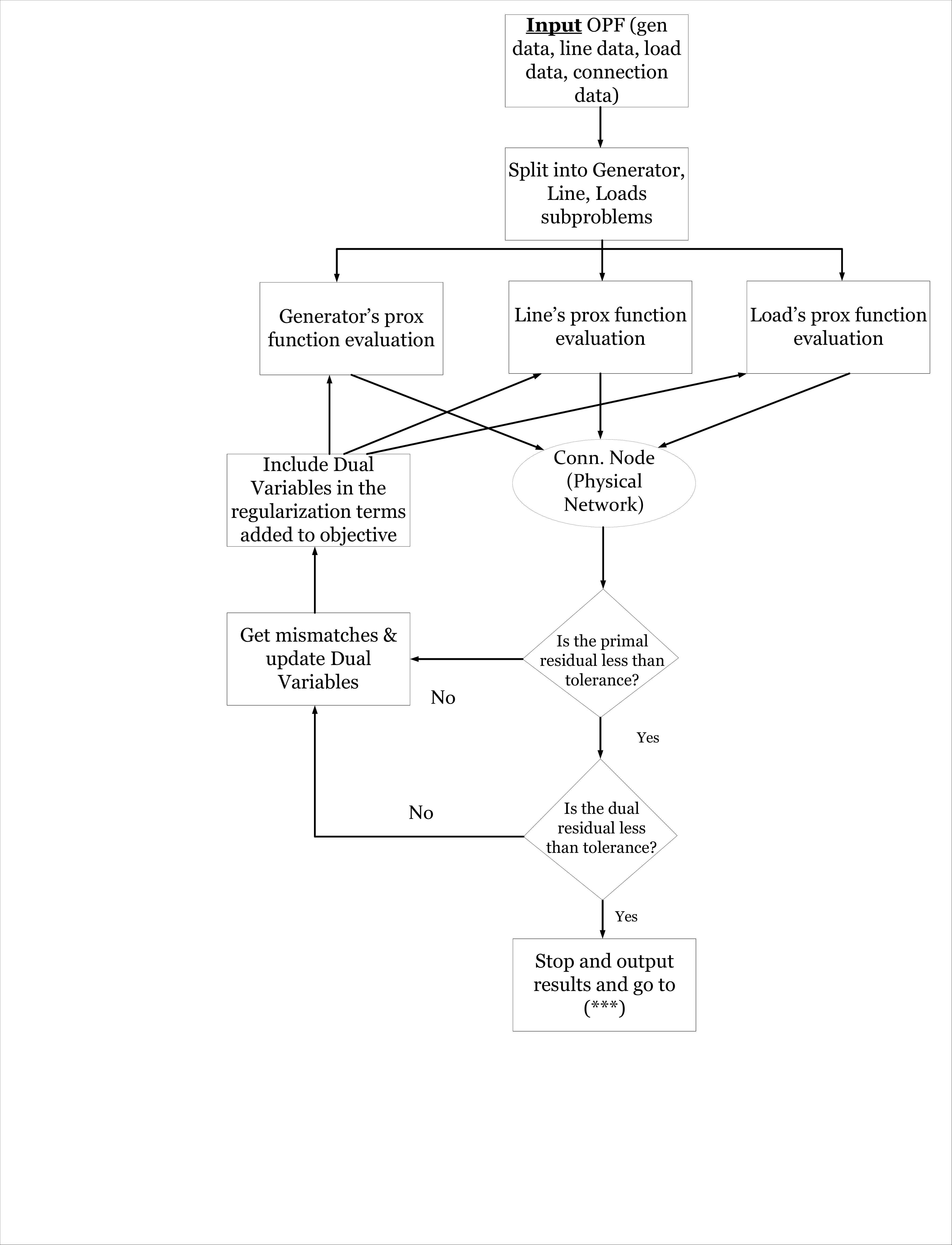}
			\caption{Flowchart for the innermost ADMM-PMP for solving OPF}
			\label{OPFADMM}
		\end{center}
	\end{figure}

We are writing here just the analytical forms already derived in \cite{KC:13}. 
	\begin{equation}\label{ADMM_Nets_Lookahead_Demand}
	\begin{array}{ll}
	\forall N_i\in\mathcal{N},\forall t_k\in\mathcal{T}\cap N_i,\forall (c)\in\mathcal{L},\tau\in\Omega\notag\\
	z_{{N_i}{t_k}}^{(c)(\tau)(\nu+1)}\\=u_{{N_i}{t_k}}^{(c)(\tau)(\nu)}+P_{{N_i}{t_k}}^{(c)(\tau)(\nu+1)}-\hat{u}_{{N_i}{t_k}}^{(c)(\tau)(\nu)}-\hat{P}_{{N_i}{t_k}}^{(c)(\tau)(\nu+1)}\\
	\xi_{{N_i}{t_k}}^{(c)(\tau)(\nu+1)}=\hat{v}_{{N_i}{t_k}}^{(c)(\tau)(\nu)}+\hat{\theta}_{{N_i}{t_k}}^{(c)(\tau)(\nu+1)}\\
	u_{{N_i}{t_k}}^{(c)(\tau)(\nu+1)}=u_{{N_i}{t_k}}^{(c)(\tau)(\nu)}+(P_{{N_i}{t_k}}^{(c)(\tau)(\nu+1)}-z_{{N_i}{t_k}}^{(c)(\tau)(\nu+1)})\\
	v_{{N_i}{t_k}}^{(c)(\tau)(\nu+1)}=v_{{N_i}{t_k}}^{(c)(\tau)(\nu)}+(\theta_{{N_i}{t_k}}^{(c)(\tau)(\nu+1)}-\xi_{{N_i}{t_k}}^{(c)(\tau)(\nu+1)})
	\end{array}
	\end{equation}
	Note that, not only do all the devices update their variables in parallel, but also, except the generators, all devices have associated with them the base-case and the contingency scenarios in each dispatch interval, each of which in turn update their respective variables in parallel as well. For the generators, the present dispatch interval and the next one are related through the ramp rate constraint ($\Delta (.)$ function). Then all the nets and the base-case/contingency scenarios associated with them update the first two set of variables in parallel and then update the next two in parallel. The proximal function for a function $g$ is given by
	\[
	\mathbf{prox}_{g,\rho}({\mathbf{v}}) = \argmin_{\mathbf{x}} (g({\mathbf{x}}) + (\rho/2)\|{\mathbf{x}} - {\mathbf{v}}\|_2^2).
	\]
	For this case, the prox messages and the Proximal Message Passing Algorithm is as follows:
	\begin{equation}\label{Prox_Lookahead_Demand}
	\begin{array}{ll}
	1.(P_{{g_q}{t_k}}^{(0)(\tau)(\nu+1)},\theta_{{g_q}{t_k}}^{(c)(\tau)(\nu+1)})\notag\\=\mathbf{prox}_{C(P)+\Delta(P),\rho}(P_{{g_q}{t_k}}^{(0)(\tau)(\nu)}-\hat{P}_{{g_q}{t_k}}^{(c)(\tau)(\nu)}-u_{{g_q}{t_k}}^{(c)(\tau)(\nu)},\\ \hat{v}_{{g_q}{t_k}}^{(c)(\tau)(\nu-1)}+\hat{\theta}_{{g_q}{t_k}}^{(c)(\tau)(\nu)}-v_{{g_q}{t_k}}^{(c)(\tau)(\nu)}),\\\forall g_q\in G,\forall(c)\in\mathcal{L},\forall\tau\in\Omega\\
	\\
	2.(P_{{T_r}{t_k}}^{(c)(\tau)(\nu+1)},\theta_{{T_r}{t_k}}^{(c)(\tau)(\nu+1)},P_{{T_r}{t_{k^{'}}}}^{(c)(\tau)(\nu+1)},\theta_{{T_r}{t_{k^{'}}}}^{(c)(\tau)(\nu+1)})\notag\\=\mathbf{prox}_{F(P)+\chi(P,\theta),\rho}(P_{{T_r}{t_k}}^{(c)(\tau)(\nu)}-\hat{P}_{{T_r}{t_k}}^{(c)(\tau)(\nu)}-u_{{T_r}{t_k}}^{(c)(\tau)(\nu)}, \\\hat{v}_{{T_r}{t_k}}^{(c)(\tau)(\nu-1)}+\hat{\theta}_{{T_r}{t_k}}^{(c)(\tau)(\nu)}-v_{{T_r}{t_k}}^{(c)(\tau)(\nu)}),
	\\\forall T_r\in \mathbb{T},\forall(c)\in\mathcal{L},\forall\tau\in\Omega\\
	\\
	3.(P_{{D_d}{t_k}}^{(c)(\tau)(\nu+1)},\theta_{{D_d}{t_k}}^{(c)(\tau)(\nu+1)})\notag\\=\mathbf{prox}_{-D,\rho}(\hat{v}_{{D_d}{t_k}}^{(c)(\tau)(\nu-1)}+\hat{\theta}_{{D_d}{t_k}}^{(c)(\tau)(\nu)}-v_{{D_d}{t_k}}^{(c)(\tau)(\nu)}),\\\forall D_d\in L,\forall(c)\in\mathcal{L},\forall\tau\in\Omega\\
	\\
	4.u_{{N_i}{t_k}}^{(c)(\tau)(\nu+1)}=u_{{N_i}{t_k}}^{(c)(\tau)(\nu)}+\hat{P}_{{N_i}{t_k}}^{(c)(\tau)(\nu+1)},\\\forall N_i\in\mathcal{N},\forall(c)\in\mathcal{L},\forall\tau\in\Omega\\
	\\
	5.v_{{N_i}{t_k}}^{(c)(\tau)(\nu+1)}=\tilde{v}_{{N_i}{t_k}}^{(c)(\tau)(\nu)}+\tilde{\theta}_{{N_i}{t_k}}^{(c)(\tau)(\nu+1)},\\\forall N_i\in\mathcal{N},\forall(c)\in\mathcal{L},\forall\tau\in\Omega
	\end{array}
	\end{equation}
	In the above, steps 1, 2, and 3 consist of the concurrent calculation of prox functions by the different devices, independently, which happens after the most recently updated values of the dual variables are ``broadcast" to the devices from the nodes, as shown in figure \ref{Broadcast}, in which, we have shown two generators ($g_1$, $g_2$), two loads ($D_1$, $D_2$), one transmission line ($T_1$), and two nodes ($N_1$, $N_2$). The direction of the arrows show the direction of conveying the most recently updated values of the dual variables.
	The calculations of steps 4 and 5, which refer to updating of the dual variables corresponding to nodal power balance and angle consistency, respectively follow the above-mentioned prox-function calculation by the devices. Once the devices finish calculating their respective prox-functions, the most recently updated decision variable values are sent back to the respective nodes, through a process called ``gather," as shown in figure \ref{Gather}
	The ADMM based PMP can be thought of as a recurring sequence of such broadcast-gather operation until the convergence criteria is satisfied, as represented in figure \ref{ADMMGen} for a generic set of several devices and nodes. The convergence criteria for the ADMM-PMP iterations is given as:\\
	\textbf{Stopping criterion for ADMM-PMP: }We can define primal and dual residuals ${\mathbf{r}}^{(\nu)}\in\mathbb{R}^{|\mathcal{N}|+|\mathcal{T}|}$ and ${\mathbf{s}}^{(\nu)}\in\mathbb{R}^{|\mathcal{N}|+|\mathcal{T}|}$, respectively (at the end of ADMM-PMP iteration $\nu$, for the
	PMP algorithm as:
	\[
	{\mathbf{r}}^{(\nu)} = \left(\hat {\mathbf{P}}^{(\nu)}, \tilde{\mathbf{\Theta}}^{(\nu)}\right),
	{\mathbf{s}}^{(\nu)} = \rho\left((({\mathbf{P}}^{(\nu)} - \mathbb{A}\times\hat {\mathbf{P}}^{(\nu)}) - ({\mathbf{P}}^{(\nu-1)} - \mathbb{A}\times\hat {\mathbf{P}}^{(\nu-1)})), (\hat {\mathbf{\Theta}}^{(\nu)}- \hat {\mathbf{\Theta}}^{(\nu-1)})\right).
	\]
	Where \begin{itemize}
		\item $\mathbb{A}\in\mathbb{R}^{|\mathcal{T}|\times|\mathcal{N}|}$ is the terminal to node incidence matrix, with 
		\begin{itemize}
			\item $\mathbb{A}(t_k,N_i)=1$ if terminal $t_k$ is connected to node $N_i$
			\item otherwise, $\mathbb{A}(t_k,N_i)=0$
		\end{itemize}
		\item ${\mathbf{P}}^{(\nu)}\in\mathbb{R}^{|\mathcal{T}|}$ is a vector of the real power iterates of all the devices-terminals (generators, transmission lines, and loads) connected to a particular node, at the end of the $\nu^{\text{-th}}$ ADMM-PMP iteration.
		\item ${\hat{\mathbf{P}}}^{(\nu)}\in\mathbb{R}^{|\mathcal{N}|}$ is a vector of the real power nodal average injections (i.e. real power iterates of all the devices (generators, transmission lines, and loads) connected to a particular node, divided by the number of devices) at the end of the $\nu^{\text{-th}}$ ADMM-PMP iteration.
		\item $\tilde{\mathbf{\Theta}}^{(\nu)}\in\mathbb{R}^{|\mathcal{T}|}$ is the difference between the voltage phase angle iterate of each device, connected to a particular node and the average voltage phase angle at the node, at the end of the $\nu^{\text{-th}}$ ADMM-PMP iteration.
		\item ${\hat{\mathbf{\Theta}}}^{(\nu)}\in\mathbb{R}^{|\mathcal{N}|}$ is a vector of the voltage phase angle nodal average (i.e. voltage phase angle iterates of all the devices (generators, transmission lines, and loads) connected to a particular node, divided by the number of devices) at the end of the $\nu^{\text{-th}}$ ADMM-PMP iteration.
	\end{itemize}
	A simple terminating criterion for prox-project message passing is when
	\[
	\| {\mathbf{r}}^{(\nu)} \|_2 \leq \epsilon^\mathrm{pri}, \qquad \|{\mathbf{s}}^{(\nu)}\|_2 \leq \epsilon^\mathrm{dual},
	\]
	where $\epsilon^\mathrm{pri}$ and $\epsilon^\mathrm{dual}$ are,
	respectively, primal and dual tolerances and the norms are second norms. The entire APMP algorithm can be pictorially represented by the three flowcharts, shown in Figures \ref{outermostAPP}, \ref{SCOPFAPP}, \ref{OPFADMM}, which respectively show the algorithmic steps for the outermost APP for attaining consensus among the beliefs about power generations about different intervals, inner APP for attaining consensus among the base-case generation value beliefs among different scenarios, and innermost ADMM-PMP, for solving the OPF for each scenario and each dispatch interval. It is to be noted, that since the algorithms are nested, that's why at the end of the inner APP iterations in Figure \ref{SCOPFAPP}, the program-flow returns to the outermost layer of Figure \ref{outermostAPP}, indicated as (**) and similarly from the end of the innermost ADMM-PMP layer of Figure \ref{OPFADMM}, to the inner APP layer of Figure \ref{SCOPFAPP}, indicated as (***).
	In the next section, we will present the numerical results pertaining to the LASCOPF simulations.

	\section{Numerical Results}
	\label{results}
\begin{figure}
		\begin{center}
			\includegraphics[height=10cm,width=20cm]{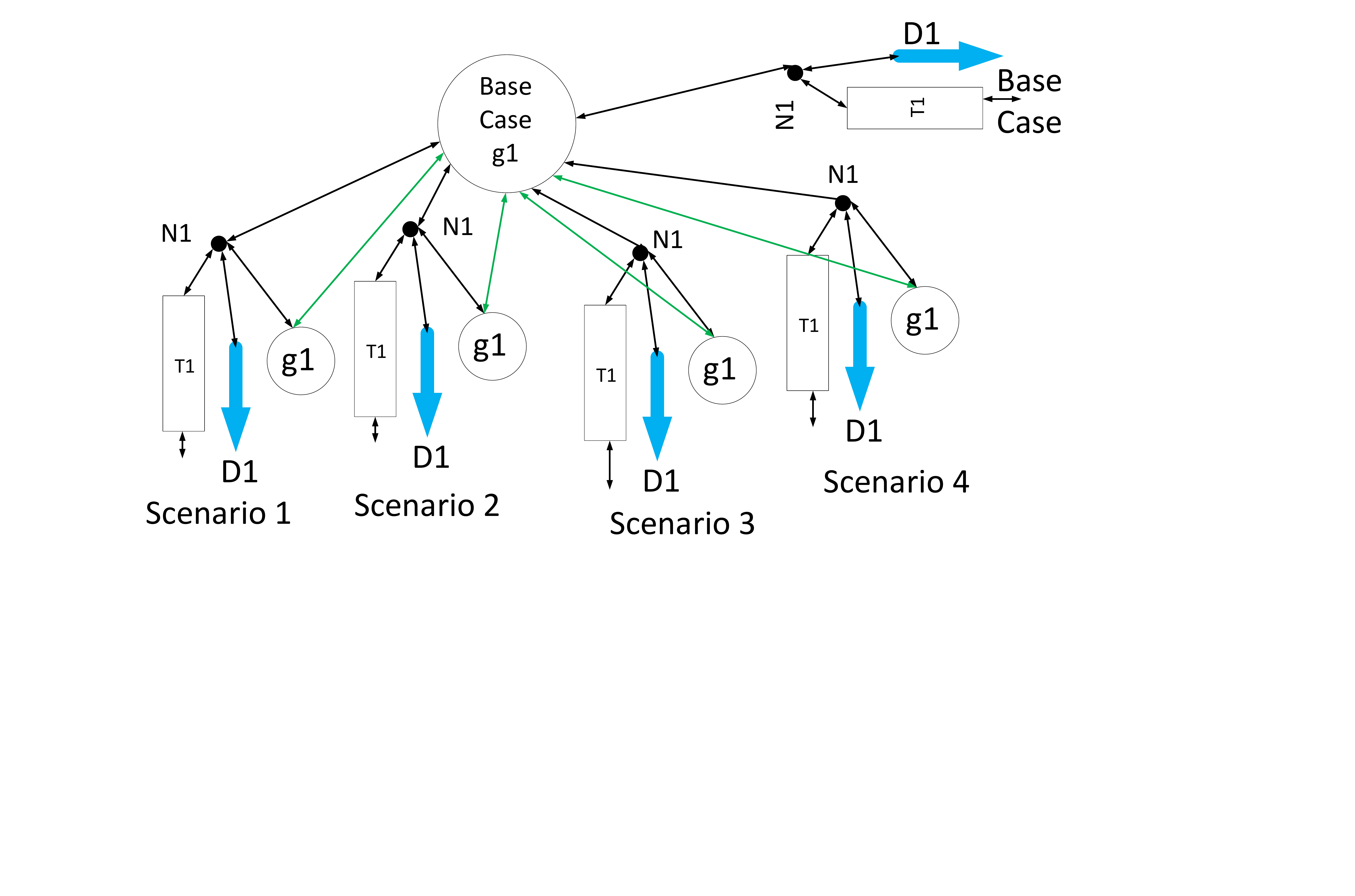}
			\vspace*{-4cm}
			\caption{APMP for SCOPF in each dispatch interval}
			\label{SCOPF_APMP}
		\end{center}
	\end{figure}

\begin{figure}
		\begin{center}
			\vspace*{-2cm}
			\includegraphics[height=10cm,width=17cm]{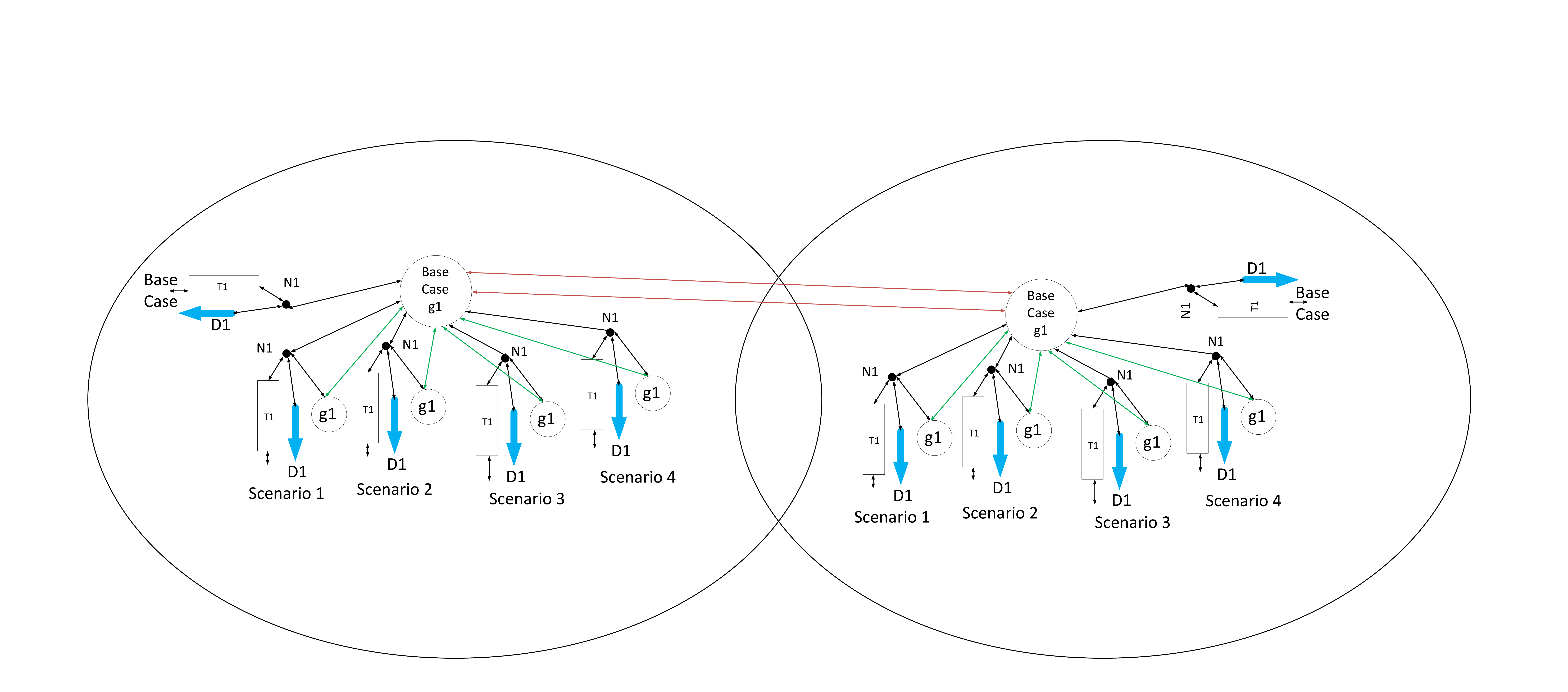}
			\caption{LASCOPF for Demand Variation with APMP Schematic for Outer APP Iterations}
			\label{LASCOPFDV_APMP}
		\end{center}
	\end{figure}

\begin{figure}
		\begin{center}
			\includegraphics[height=17.5cm,width=21cm, angle=90]{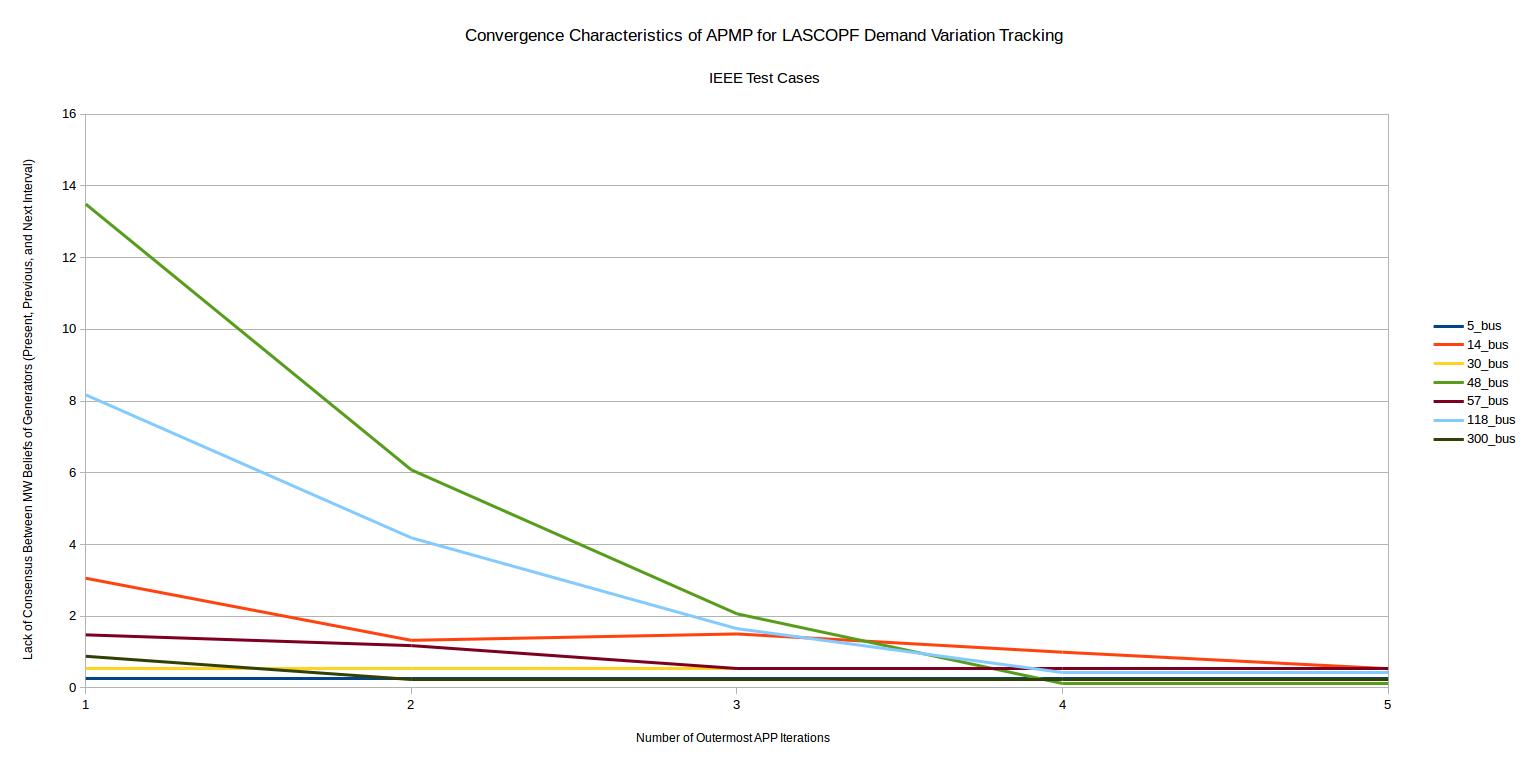} 
			\caption{Convergence Characteristics of IEEE Test Systems}
			\label{SummaryConv}
		\end{center}
	\end{figure}

	In this section, we present the results of the APMP algorithm for the LASCOPF problem instance for tracking demand variation for the IEEE Test cases \cite{WashIEEE} and for a 5 bus system, with data shown in tables \ref{table:5BusGen} and \ref{table:5BusLine}.
	\begin{table}[ht] 
		
		\caption{Generator Data for 5 Bus System.} 
		
		\centering 
		
		\begin{tabular}{| c | c | c | c | c | c | c | c | c |} 
			
			\hline\hline 
			
			Node & A (\$/$\text{MWh}^2$) & B(\$/MWh) & C(\$/h)  & $\overline{P}$ & $\underline{P}$ &  $\overline{R}$ & $\underline{R}$ & Sched. MW \\ [0.5ex] 
			
			
			\hline 
			
			1  &	0.0430293  &	20  &	0  &	332.4  &	0  &	20  &	-20  &	140.765 \\ [0.5ex] 
			\hline
			2  &	0.25  &	20  &	0  &	140  &	0  &	15  &	-15  &	24.2275 \\ [0.5ex] 
			\hline
		\end{tabular} 
		
		\label{table:5BusGen} 
		
	\end{table}
	\begin{table}[ht] 
		
		\caption{Line Data for 5 Bus System.} 
		
		\centering 
		
		\begin{tabular}{| c | c | c | c | c |} 
			
			\hline\hline 
			
			From Node & To Node & Resistance (pu) & Reactance (pu)  & Flow Limit (MW) \\ [0.5ex] 
			
			
			\hline 
			1  &	2  &	0.02  &	0.06  &		100 \\ [0.5ex] 
			\hline
			1  &	3  &	0.08  &	0.24  &		100 \\ [0.5ex] 
			\hline
			2  &	3  &	0.06  &	0.18  &		100 \\ [0.5ex] 
			\hline
			2  &	4  &	0.06  &	0.18  &		100 \\ [0.5ex] 
			\hline
			2  &	5  &	0.04  &	0.12  &		100 \\ [0.5ex] 
			\hline
			3  &	4  &	0.01  &	0.03  &		100 \\ [0.5ex] 
			\hline
			4  &	5  &	0.08  &	0.24  &		100 \\ [0.5ex] 
			\hline
		\end{tabular} 
		
		\label{table:5BusLine} 
		
	\end{table}
	In table \ref{table:5BusGen}, A, B, and C are the quadratic, linear, and no load coefficients of the cost curve, respectively. $\overline{P}$, $\underline{P}$, $\overline{R}$, and $\underline{R}$ are respectively, the maximum, minimum generating limits in MW and maximum and minimum ramping limits expressed in MW/dispatch interval. While performing the actual simulations, we have observed that it is very hard to solve the $(N-1)$ SCOPFs just by using ADMM-PMP, perhaps due to difficulty in choosing the appropriate tuning parameter values. So, we have used a version of APMP for solving the SCOPFs as well, as illustrated in figure \ref{SCOPF_APMP}. 
	Here, we apply the ADMM-PMP to solve OPFs for the base case as well as each contingency scenario separately and use an outer layer of APP over all the scenarios to reach a consensus regarding the pre-contingency dispatch (the message exchanges regarding the beliefs of each scenario for the value of dispatch, to reach consensus, are shown with the green lines). We have been successful at solving the SCOPFs using this approach for the IEEE 5, 14, 30, 48, 57, 118, and 300 bus test cases with different lines marked for contingency, and also were able to identify those problem instances that are infeasible. Thereafter, for solving the LASCOPF, we can now treat the SCOPF solver as one unit, and as figure \ref{LASCOPFDV_APMP} shows, use the outermost APP layer to exchange messages about the beliefs of the different interval-wise generation values to reach a consensus. 
	All the other LASCOPF problems can be solved in the exact same manner. We have run all these simulations on a Dell Inspiron 17R laptop computer powered by a 4x Intel(R) Core(TM) i7-4500U CPU running at 1.80 GHz, with a RAM of capacity 8054 MB and the OS is Ubuntu 16.04.4 LTS. We have coded all the simulation programs in C++11, with the generator optimization solvers being implemented in two modes: 
	\begin{itemize}
		\item CVXGEN custom solvers \cite{cvxgen01}, \cite{cvxgen02}, \cite{cvxgen03}, \cite{cvxgen04} for generators' optimization for fully distributed APMP (APP+ADMM-PMP)
		\item GUROBI solvers for generators' optimization for fully distributed APMP (APP+ADMM-PMP)
	\end{itemize} 
	The compiler used is GCC version 4.8.4. In the above-mentioned two methods, only the solvers are different. However, the implementation is a combination of two outer layers of coarse grained APP and one innermost layer of fine-grained ADMM-PMP. We have observed no or minimal disagreement between the results obtained by using these two solvers. However, the time taken by GUROBI is much longer than that by CVXGEN custom solvers \cite{cvxgen01}, \cite{cvxgen02}, \cite{cvxgen03}, \cite{cvxgen04}. Hence, we will only present the results we obtained using CVXGEN. We have experimented with IEEE test cases with 5, 14, 30, 48, 57, 118, and 300 bus systems, with contingencies of some or all lines analyzed for contingencies, and for several of dispatch intervals. The simulation software that we have designed can determine (based on a combination logic of the number of iterations and convergence trend of the residuals of the different stages of the algorithm, as well as the different parts of the system) whether a particular problem instance is solvable or feasible or not. Out of those cases that we found out to be feasible, we will present here the results of the following cases:\\
	\begin{itemize}
		\item 5 bus system with each line analyzed for contingency
		\item 14 bus system with each line analyzed for contingency
		\item 30 bus system with lines (from node-to node) (5-7), (6-28), (10-21), and (12-16) analyzed for contingency
		\item 48 bus system with lines (1-5), (2-6), and (3-24) analyzed for contingency
		\item 57 bus system with lines (9-12), (12-17), (38-48), and (24-26) analyzed for contingency
		\item 118 bus system with lines (1-3) and (2-12) analyzed for contingency
		\item 300 bus system with lines (7-131) and (201-204) analyzed for contingency
	\end{itemize}
	Table \ref{table:ConvergenceChar} below lists the simulation metrics of the above-mentioned cases and the ST is the simulation time in seconds that it takes to solve the respective problems, if we implement a fully and complete nested parallelization. The numbers there represent the residual of the outermost APP layer of the algorithm, which is the norm-2 of the disagreement between the different generation MW beliefs regarding the present, previous, and next interval generation values. 
	\begin{table}[ht] 
		
		\caption{Convergence Metrics for the IEEE Test Cases} 
		
		\centering 
		
		\begin{tabular}{| c | c | c | c | c | c | c | c |} 
			
			\hline\hline 
			Iter &	5 Bus &	14 Bus &	30 Bus &	48 Bus &	57 Bus &	118 Bus &	300 Bus \\ [0.5ex] 
			
			\hline 
			
			1 &	0.2673 &	3.05719 &	0.537591 &	13.493 &	1.47779 &	8.17101 &	0.8805 \\ [0.5ex]
			\hline
			2 &	**0.2673 &	1.32581 &	**0.537591 &	6.0792 &	1.17792 &	4.18356 &	0.2317 \\ [0.5ex]
			\hline
			3 &	**0.2673 &	1.50355 &	**0.537591 &	2.06586 &	0.526238 &	1.6507 &	**0.2317 \\ [0.5ex]
			\hline
			4 &	**0.2673 &	0.993954 &	**0.537591 &	0.111532 &	**0.526238 &	0.431 &	**0.2317 \\ [0.5ex]
			\hline
			5 &	**0.2673 &	0.534623 &	**0.537591 &	**0.111532 &	**0.526238 &	**0.431 &	**0.2317 \\ [0.5ex]
			\hline
			\hline
			ST (s) &	0.676335 &	205.667 &	1.28923 &	79.163 &	162.108 &	848.727 &	918.707 \\ [0.5ex]
			\hline
		\end{tabular} 
		
		\label{table:ConvergenceChar} 
		
	\end{table}
	The ** sign next to the residuals mean that the outer iteration has already converged by the corresponding iteration count and we have fixed the subsequent residual values, just to help us draw the plot, that appears in fig \ref{SummaryConv}. 
	\begin{figure}
		\begin{center}
			\includegraphics[height=17.5cm,width=21cm, angle=90]{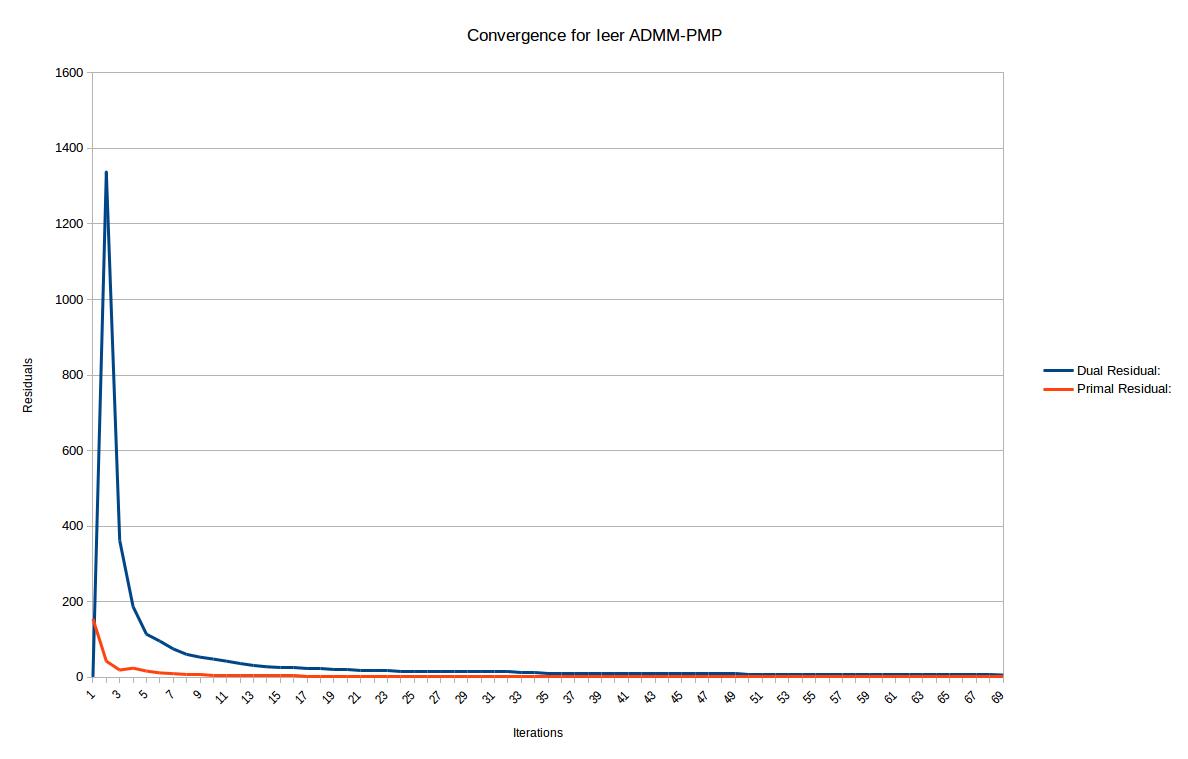} 
			\caption{Convergence Characteristics of Innermost ADMM-PMP for 300 bus case (contingency scenario:2, dispatch interval:3)}
			\label{ADMM-PMP300}
		\end{center}
	\end{figure}
	The load variation data for the 5 bus and 14 bus cases over 5 dispatch intervals, that we have simulated for, are explicitly shown in tables \ref{table:5LASCOPFLoadModified} and \ref{table:14LASCOPFLoadModified} respectively. The ones for the 30, 48, 57, 118, and 300 bus systems are shown as only the variations from the base case value (that appear on the UW IEEE test case archive https://labs.ece.uw.edu/pstca/) of the particular loads in particular intervals in tables \ref{table:30LASCOPFLoadModified}, \ref{table:48LASCOPFLoadModified}, \ref{table:57LASCOPFLoadModified}, \ref{table:118LASCOPFLoadModified}, and \ref{table:300LASCOPFLoadModified} respectively. The base case values occur in the first interval.
	For the innermost OPF simulations, which we solve by using ADMM-PMP, we have used the primal and dual residual tolerances to be 0.06 and 0.6, respectively (the reason we use a higher value for the dual residual is that we have observed that we get accurate enough solution for even somewhat higher value of the dual residual) and we have used a discrete version of proportional+derivative controller to adjust the value of $\rho$ for the first 3000 iterations, such that at each iteration the relationship $\rho\times\epsilon_{primal}=\epsilon_{dual}$ is maintained, where $\epsilon_{primal}$ and $\epsilon_{dual}$ are respectively the primal and dual residuals. After the first 3000 iterations, if the algorithm hasn't still converged, then $\rho$ is held fixed at the last value. The initial value of $\rho$ at the beginning of the iterations is taken as 1. We have shown in Figure \ref{ADMM-PMP300}, one typical plot for the first 70 iterations of the innermost ADMM-PMP iterations for the 300 bus case for contingency scenario:2 and dispatch interval:3. This set of ADMM-PMP iterations was executed for the last iteration of the outer APP layers, both for consensus among scenarios, as well as, among dispatch intervals. The following are the APP parameters for outer iterations for attaining consensus among the different base/contingency scenarios for solving SCOPFs.
	\begin{itemize}
		\item $\alpha_{\text{SCOPF}}=5$ for $\nu\leq5$, $\alpha_{\text{SCOPF}}=3$ for $5<\nu\leq10$, $\alpha_{\text{SCOPF}}=2.5$ for $10<\nu\leq15$, $\alpha_{\text{SCOPF}}=1.25$ for $15<\nu\leq20$, and $\alpha_{\text{SCOPF}}=0.5$ for $\nu>20$, $\beta_{\text{SCOPF}}=200$, $\gamma_{\text{SCOPF}}=100$ ($\nu$ is the APP iteration count)
		\item Final Tolerance $\epsilon_{\text{SCOPF}}$: 0.7
	\end{itemize}
	Following are the APP parameters for outermost iterations for attaining consensus among the different MW outputs in different intervals, limited by ramp-rate constraints.
	\begin{itemize}
		\item $\alpha_{\text{LASCOPF}}=10$ for $\mu_{APP}\leq5$, $\alpha_{\text{LASCOPF}}=5$ for $5<\mu_{APP}\leq10$, $\alpha_{\text{LASCOPF}}=2.5$ for $10<\mu_{APP}\leq15$, $\alpha_{\text{LASCOPF}}=1.25$ for $15<\mu_{APP}\leq20$, and $\alpha_{\text{LASCOPF}}=0.5$ for $\mu_{APP}>20$, $\beta_{\text{LASCOPF}}=200$, $\gamma_{\text{LASCOPF}}=100$ ($\mu_{APP}$ is the outermost APP iteration count)
		\item Final Tolerance $\epsilon_{\text{LASCOPF}}$: 0.6
	\end{itemize}
	As can be seen, for both the inner and outer APP iterations, we have used changing step-length. The way we chose those values are ad-hoc and gives a reasonable balance between accuracy and solve-time.
	\begin{table}[ht] 
		
		\caption{Variation of Load for 5 Bus System.} 
		
		\centering 
		
		\begin{tabular}{| c | c | c | c | c | c |} 
			
			\hline\hline 
			
			Conn. Node & MW, Int.-1 & MW, In.-2 & MW, Int.-3  & MW, Int.-4 & MW, Int.-5 \\ [0.5ex] 
			
			
			\hline 
			
			2 &	20 &	30 &	20 &	20 &	20 \\ 
			\hline
			3 &	45 &	40 &	43 &	43 &	43 \\ 
			\hline
			4 &	40 &	40 &	45 &	45 &	45 \\ 
			\hline
			5 &	60 &	65 &	65 &	65 &	65 \\ 
			\hline
		\end{tabular} 
		
		\label{table:5LASCOPFLoadModified} 
		
	\end{table}
	\begin{table}[ht] 
		
		\caption{Variation of Load for 14 Bus System.} 
		
		\centering 
		
		\begin{tabular}{| c | c | c | c | c | c |} 
			
			\hline\hline 
			
			Conn. Node & MW, Int.-1 & MW, In.-2 & MW, Int.-3  & MW, Int.-4 & MW, Int.-5 \\ [0.5ex] 
			
			
			\hline 
			
			2 &	21.7 &	16.7 &	26.7 &	24.2 &	14.2 \\ 
			\hline
			3 &	94.2 &	89.2 &	99.2 &	96.7 &	86.7 \\ 
			\hline
			4 &	47.8 &	42.8 &	52.8 &	50.3 &	40.3 \\ 
			\hline
			5 &	7.6 &	2.6 &	12.6 &	10.1 &	0.1 \\ 
			\hline
			6 &	11.2 &	6.2 &	16.2 &	13.7 &	3.7 \\ 
			\hline
			9 &	29.5 &	24.5 &	34.5 &	32 &	22 \\ 
			\hline
			10 &	9 &	4 &	14 &	11.5 &	1.5 \\ 
			\hline
			11 &	3.5 &	1.5 &	8.5 &	6 &	4 \\ 
			\hline
			12 &	6.1 &	1.1 &	11.1 &	8.6 &	1.4 \\ 
			\hline
			13 &	13.5 &	8.5 &	18.5 &	16 &	6 \\ 
			\hline
			14 &	14.9 &	9.9 &	19.9 &	17.4 &	7.4 \\ 
			\hline
		\end{tabular} 
		
		\label{table:14LASCOPFLoadModified} 
		
	\end{table}
	\begin{table}[ht] 
		
		\caption{Variation of Load for 30 Bus System.} 
		
		\centering 
		
		\begin{tabular}{| c | c | c | c | c |} 
			
			\hline\hline 
			
			Conn. Node & MW, Int.-2 & MW, In.-3 & MW, Int.-4  & MW, Int.-5 \\ [0.5ex] 
			
			
			\hline 
			
			10 &	0 &	0 &	10 &	0 \\ 
			\hline
			12 &	5 &	0 &	0 &	0 \\ 
			\hline
			18 &	0 &	3 &	0 &	0 \\ 
			\hline
			24 &	0 &	0 &	0 &	12 \\ 
			\hline
		\end{tabular} 
		
		\label{table:30LASCOPFLoadModified} 
		
	\end{table}
	\begin{table}[ht] 
		
		\caption{Variation of Load for 48 Bus System.} 
		
		\centering 
		
		\begin{tabular}{| c | c | c | c | c |} 
			
			\hline\hline 
			
			Conn. Node & MW, Int.-2 & MW, In.-3 & MW, Int.-4  & MW, Int.-5 \\ [0.5ex] 
			
			
			\hline 
			
			5 &	0 &	100 &	0 &	0 \\ 
			\hline
			7 &	0 &	0 &	0 &	100 \\ 
			\hline
			19 &	0 &	22 &	0 &	0 \\ 
			\hline
			20 &	40 &	0 &	0 &	0 \\ 
			\hline
			27 &	0 &	0 &	0 &	-20 \\ 
			\hline
			29 &	0 &	0 &	-47 &	0 \\ 
			\hline
			32 &	20 &	0 &	-14 &	0 \\ 
			\hline
			38 &	0 &	-100 &	0 &	0 \\ 
			\hline
			44 &	0 &	0 &	100 &	0 \\ 
			\hline
		\end{tabular} 
		
		\label{table:48LASCOPFLoadModified} 
		
	\end{table}
	\begin{table}[ht] 
		
		\caption{Variation of Load for 57 Bus System.} 
		
		\centering 
		
		\begin{tabular}{| c | c | c | c | c |} 
			
			\hline\hline 
			
			Conn. Node & MW, Int.-2 & MW, In.-3 & MW, Int.-4  & MW, Int.-5 \\ [0.5ex] 
			
			
			\hline 
			
			8 &	5 &	0 &	0 &	0 \\ 
			\hline
			10 &	0 &	0 &	0 &	5 \\ 
			\hline
			14 &	0 &	3 &	0 &	0 \\ 
			\hline
			16 &	0 &	0 &	-20 &	0 \\ 
			\hline
			25 &	0.4 &	0 &	0 &	0 \\ 
			\hline
			29 &	0 &	0 &	30 &	0 \\ 
			\hline
			32 &	0 &	1 &	0 &	0 \\ 
			\hline
			41 &	0 &	3.7 &	0 &	0 \\ 
			\hline
			47 &	0 &	6 &	0 &	0 \\ 
			\hline
		\end{tabular} 
		
		\label{table:57LASCOPFLoadModified} 
		
	\end{table}
	\begin{table}[ht] 
		
		\caption{Variation of Load for 118 Bus System.} 
		
		\centering 
		
		\begin{tabular}{| c | c | c | c | c |} 
			
			\hline\hline 
			
			Conn. Node & MW, Int.-2 & MW, In.-3 & MW, Int.-4  & MW, Int.-5 \\ [0.5ex] 
			
			
			\hline 
			
			36 &	0 &	0 &	20 &	0 \\ 
			\hline
			51 &	-20 &	0 &	0 &	0 \\ 
			\hline
			70 &	0 &	-10 &	0 &	0 \\ 
			\hline
			77 &	0 &	0 &	0 &	40 \\ 
			\hline
		\end{tabular} 
		
		\label{table:118LASCOPFLoadModified} 
		
	\end{table}
	\begin{table}[ht] 
		
		\caption{Variation of Load for 300 Bus System.} 
		
		\centering 
		
		\begin{tabular}{| c | c | c | c | c |} 
			
			\hline\hline 
			
			Conn. Node & MW, Int.-2 & MW, In.-3 & MW, Int.-4  & MW, Int.-5 \\ [0.5ex] 
			
			
			\hline 
			
			1 &	1 &	0 &	0 &	0 \\ 
			\hline
			15 &	0 &	4 &	0 &	0 \\ 
			\hline
			145 &	1 &	0 &	0 &	0 \\ 
			\hline
			209 &	0 &	0 &	12 &	0 \\ 
			\hline
			217 &	0 &	0 &	0 &	20 \\ 
			\hline
			9533 &	1 &	0 &	0 &	0 \\ 
			
			\hline
		\end{tabular} 
		
		\label{table:300LASCOPFLoadModified} 
		
	\end{table}
	Figure \ref{SummaryConv} shows the outer APP convergence for this example problem.
	As future research, we would like to explore proven means of tuning the path length for minimum number of iterations to converge.
	\section{Conclusion}
	\label{conclusion}
	In this paper, we have presented the Look-Ahead SCOPF (LASCOPF) considering variation of load demand over several dispatch time intervals and ability of the LASCOPF model to successfully track the temporal demand variation. We have also presented a completely decentralized (i.e. without any central coordinator) computational algorithm to implement the calculations for such a massive scale problem. Through the simulation examples, we have demonstrated the scalability of the algorithm and the effectiveness of the methodology for dispatching systems where load varies over time. The next step in the research in this direction will be to represent the post-contingency states corresponding to each contingency scenario and the system restoration to security over multiple dispatch intervals. We will be presenting that topic in our next paper.
	
	\appendix
	
	\section{Acknowledgment}
	
	\noindent The authors were supported, in part, by the National Science Foundation
	under grant ECCS-1406894. The authors would like to thank Mahdi Kefayati, PhD graduate (2014), Department of Electrical and Computer Engineering, The University of Texas at Austin, Patryk Radyjowski, PhD student, Department of Mechanical Engineering, The University of Texas at Austin, Matt Kraning, Eric Chu, PhD graduates (2013) from Stanford University, and Prof. Stephen Boyd from Stanford University for help with software and mathematical modeling related issues.

	\bibliographystyle{elsarticle-num}
	\bibliography{Look_Ahead_Demand_ADMM_journal.bib}
\end{document}